\documentclass[a4paper, 12pt]{iopart}

\usepackage[utf8]{inputenc}
\usepackage[T1]{fontenc}
\usepackage{ae}
\usepackage[automark]{scrpage2}
\usepackage{array}
\usepackage{amsfonts, amssymb, amsthm, amstext}
\usepackage{graphicx}
\usepackage{hyperref}
\usepackage{cite}

\usepackage[usenames, dvipsnames]{color}

\newcommand{\prespectrometer}{pre-spec\-tro\-meter}

\begin{document}

\title{Stochastic Heating by ECR as a Novel Means of Background Reduction in the KATRIN Spectrometers}

\author{S. Mertens$^1$, A. Beglarian$^1$, L. Bornschein$^1$, G. Drexlin$^1$, F.M. Fr{\"a}nkle$^{1,2}$, D. Furse$^3$, F. Gl{\"u}ck$^{1,4}$, S. G{\"o}rhardt$^1$, O. Kr{\"o}mer$^1$, B. Leiber$^1$, K. Schl{\"o}sser$^1$, T. Th{\"u}mmler$^1$, N. Wandkowsky$^1$, S. W{\"u}stling$^1$}

\address{1 KCETA, Karlsruhe Institute of Technology, Karlsruhe, Germany}
\address{2 Department of Physics, University of North Carolina, Chapel Hill, NC, USA}
\address{3 Laboratory of Nuclear Science, Massachusetts Institute of Technology, Cambridge, MA, USA}
\address{4 Research Institute for Nuclear and Particle Physics, Theory Dep., Budapest, Hungary}

\ead{susanne.mertens@kit.edu}


\begin{abstract}
The primary objective of the KATRIN experiment is to probe the absolute neutrino mass scale with a sensitivity of 200~meV (90\% C.L.) by precision spectroscopy of tritium $\beta$-decay. To achieve this, a low background of the order of $10^{-2}$~cps in the region of the tritium $\beta$-decay endpoint is required. 
Measurements with an electrostatic retarding spectrometer have revealed that electrons, arising from nuclear decays in the volume of the spectrometer, are stored over long time periods and thereby act as a major source of background exceeding this limit. 
In this paper we present a novel active background reduction method based on stochastic heating of stored electrons by the well-known process of electron cyclotron resonance (ECR). A successful proof-of-principle of the ECR technique was demonstrated in test measurements at the KATRIN \prespectrometer{}, yielding a large reduction of the background rate. In addition, we have carried out extensive Monte Carlo simulations to reveal the potential of the ECR technique to remove all trapped electrons in a few ms with negligible loss of measurement time in the main spectrometer. This would allow the KATRIN experiment attaining its full physics potential.
\end{abstract}

\section{Introduction}
The Karlsruhe Tritium Neutrino (KATRIN) experiment~\cite{DesignReport} is a next generation direct neutrino mass experiment, currently under construction at Tritium Laboratory Karlsruhe at the Karlsruhe Institute of Technology (KIT) Campus North site. KATRIN is designed to measure directly the effective electron anti-neutrino mass $\text{m}_{\overline{\nu}_e}$, defined as  
\begin{equation}
  \text{m}_{\overline{\nu}_e} = \sqrt{\sum\limits_{i=1}^{3}|\text{U}_{\text{e}i}|^2\cdot \text{m}_i^2}, 
\end{equation}
where $\text{U}_{\text{e}i}$ denotes elements of the Pontecorvo-Maki-Nakagawa-Sakata leptonic mixing matrix and $\text{m}_i$ denote the neutrino mass eigenstates~\cite{OttenWeinheimer}. The KATRIN design sensitivity of 200~meV at 90\% confidence level will allow to cover the quasi-degenerate neutrino mass pattern (where $\text{m}_1\approx\text{m}_2\approx\text{m}_3$) and to investigate the role of relic neutrinos from the Big Bang in the evolution of large-scale structures in the universe~\cite{DesignReport}. 

\subsection{The KATRIN setup}
The experiment analyzes the shape of the tritium $\beta$-decay spectrum in a narrow region close to its endpoint at $E_0 = 18.6$~keV. A non-zero neutrino mass reduces the maximum energy of the electron and changes the shape of the tritium $\beta$-spectrum in the immediate vicinity of the endpoint. To reach the neutrino mass sensitivity a high energy resolution spectrometer, a high signal count rate as well as a low background rate are required.

\begin{figure}[]
\begin{center}
\includegraphics[width = \textwidth]{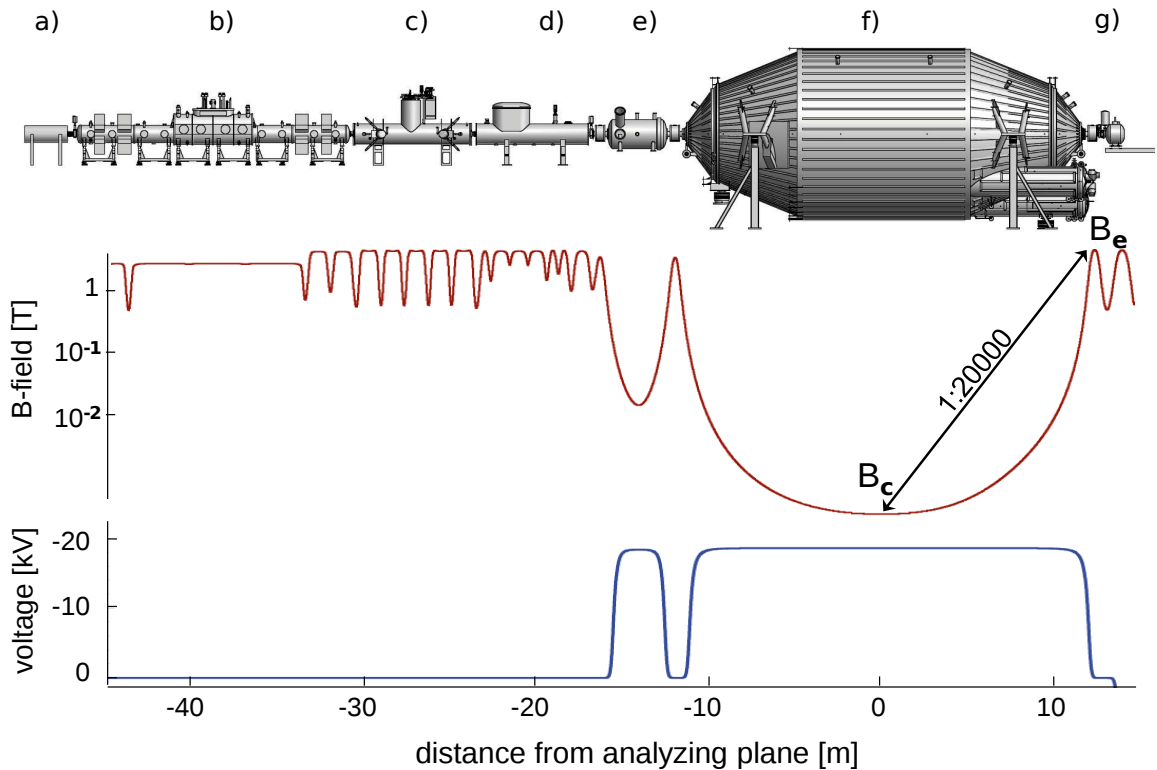}
\caption{Main components of the KATRIN experimental setup. a: Rear section, b: windowless gaseous tritium source, c: differential pumping section, d: cryogenic pumping section, e: \prespectrometer{}, f: main spectrometer, g: focal plane detector. Below, the magnetic field and the electric potential along the beam axis are displayed. In both spectrometers the MAC-E filter principle is applied: As the electric potential is increased to $\text{U}_{\text{ret}}=-18.6$~kV to filter the $\beta$-electrons, the magnetic field drops from $\text{B}_{\text{e}}=6$~T to $\text{B}_{\text{c}}=3\cdot10^{-4}$~T. The magnetic flux of $\Phi=191~\text{T}\text{cm}^2$ is conserved along the beam line.}
\label{fig:KATRIN}
\end{center}
\end{figure}

The 70~m long KATRIN setup (see figure~\ref{fig:KATRIN}) combines a molecular windowless gaseous tritium source (WGTS) of highest stability and luminosity~\cite{WGTS} with a high resolution electrostatic spectrometer for precision $\beta$-spectroscopy. A magnetic guidance system directs the electrons created in the tritium source to the spectrometer section, consisting of a smaller \prespectrometer{} providing the option to filter out low-energy electrons, and a larger main spectrometer for precision energy analysis. Both spectrometers work as electrostatic filters and transmit only those electrons that have sufficient kinetic energy to pass the retarding potential. The transmitted electrons are counted at a segmented silicon (Si-PIN) detector. By measuring the count rate for different filter voltages, the shape of the integral energy spectrum close to the tritium endpoint can be determined and $\text{m}_{\overline{\nu}_e}$ can be deduced.

To collimate the momenta of the $\beta$-electrons created isotropically in the WGTS, the magnetic field drops by four orders of magnitude from the entrance/exit of the spectrometer to its center, the so called analyzing plane. The combination of electrostatic filtering with magnetic adiabatic collimation is called the MAC-E-Filter principle~\cite{MACE1, MACE2} and allows for both large solid angle acceptance and high energy resolution. 

\subsection{Background due to stored electrons}
\begin{figure}
\begin{center}
\includegraphics[width = \textwidth]{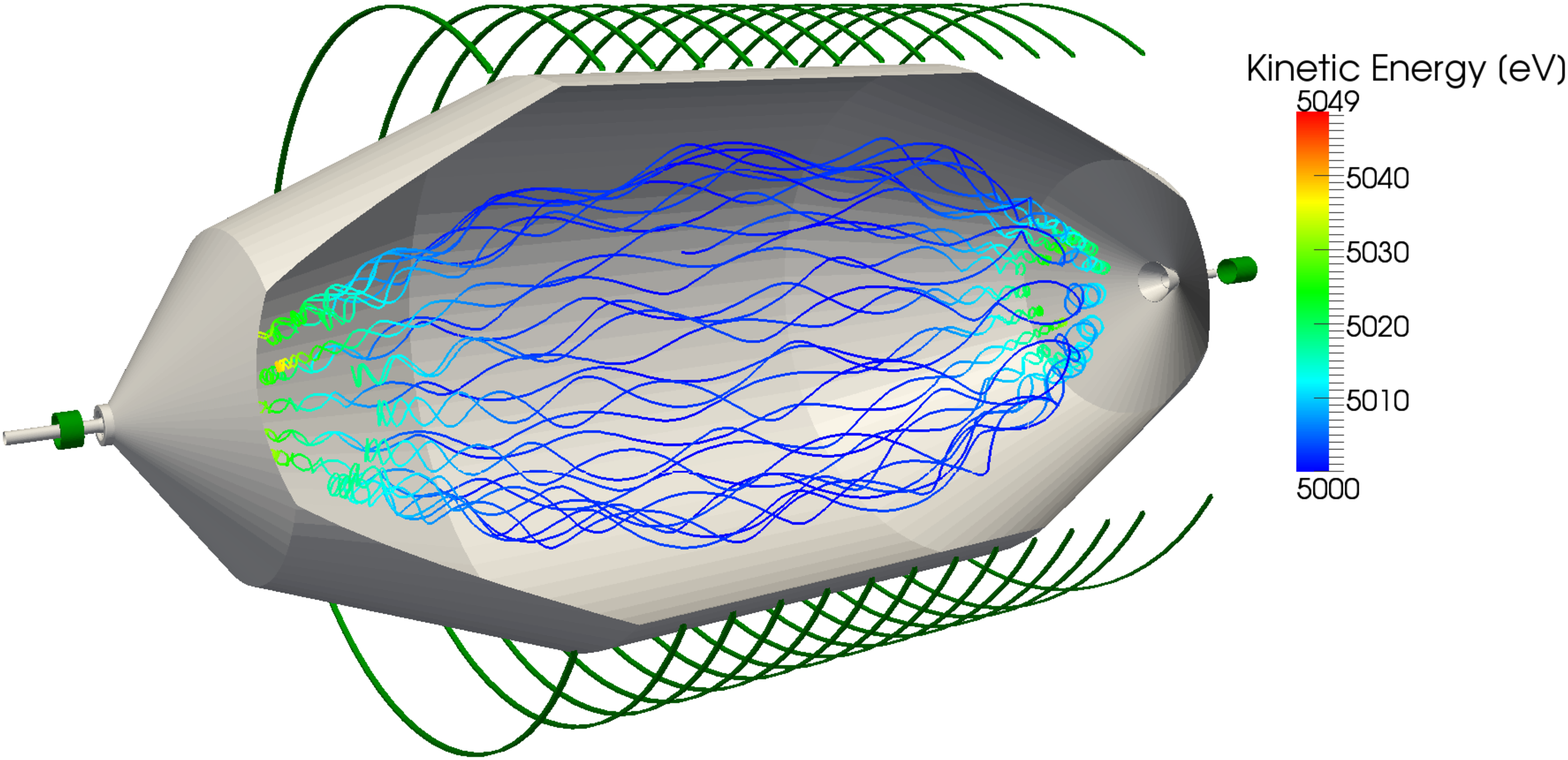}
\includegraphics[width = \textwidth]{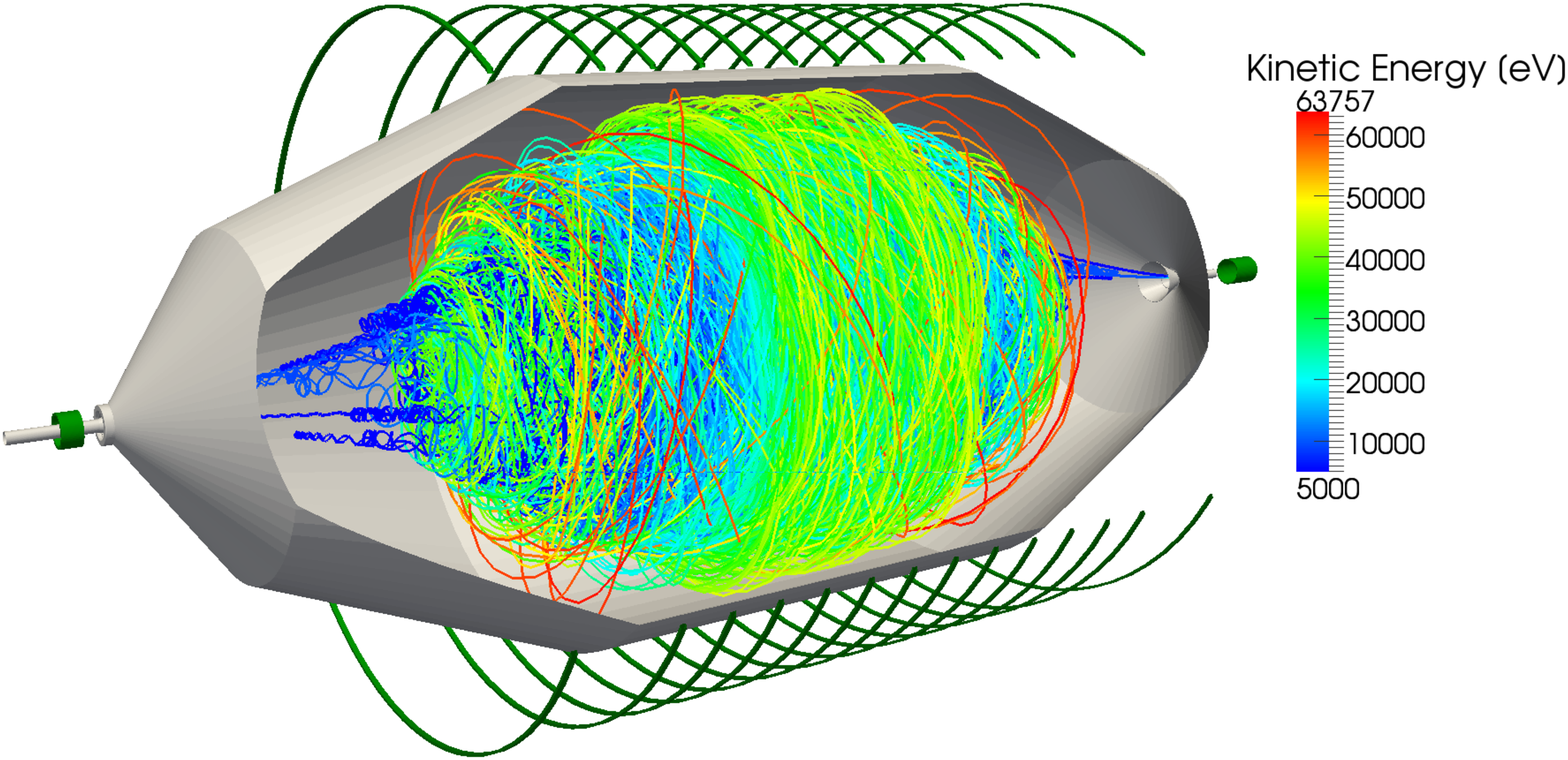}
\caption{The top picture shows a 5~keV electron under stable storage conditions in the KATRIN main spectrometer over a time period of $2\cdot10^{-5}$~s. The electron trajectory is a superposition of the fast cyclotron motion, an axial oscillation, and a much slower magnetron drift. In the bottom picture an electron trajectory in the presence of an RF-field is shown. After about 1~ms, the electron's cyclotron radius becomes larger than the radius of the spectrometer and the electron is absorbed. The color coding denotes the kinetic energy (note that the energy scale in each picture is different).}
\label{fig:trapped_radon}
\end{center}
\end{figure}

The magnetic field configuration of a MAC-E-Filter generically creates a magnetic bottle for light charged particles being produced in the volume of the spectrometer. When propagating towards the entrance or exit parts of the spectrometer, they are magnetically mirrored by the increasing magnetic field (see figure~\ref{fig:trapped_radon})~\cite{MagneticMirror1, MagneticMirror2, MagneticMirrorBook}. Sources of stored electrons with energies up to several hundred keV stem from $\alpha$-decays of the radon isotopes ${}^{219,220}$Rn and $\beta$-decays of tritium molecules (HT, $\text{T}_2$) occurring in the volume of the spectrometer. Stored electrons in the multi-keV-range mainly lose energy via ionization of residual gas molecules thereby generating secondary electrons at lower energies~\cite{scattering1, scattering2, scattering3}. Due to the excellent ultra-high vacuum (UHV) conditions~\cite{Vacuum, NEG} at the KATRIN spectrometers, the storage time of the keV-range electrons can last up to several hours. In this time period hundreds of secondary electrons are produced.

Due to their low energy, secondary electrons are eventually released from the magnetic trap. Accelerated by the retarding potential, they hit the Si-PIN detector and thus produce background in the narrow energy interval of the signal $\beta$-decay electrons~\cite{Susanne}. 

Background bunches due to stored electrons from tritium beta decays were already observed in the Troitsk neutrino mass experiment~\cite{Belesev}. Furthermore, measurements at the \prespectrometer{}~\cite{Fraenkle} and corresponding Monte Carlo (MC) simulations~\cite{Nancy} revealed that at a pressure level of $p~=~10^{-11}$~mbar a single nuclear decay can lead to an enhanced level of background of more than $3\cdot10^{-2}$~cps for several hours, which is well above the design limit of $10^{-2}$~cps. Of even greater importance is the fact that this background shows large non-Poissonian fluctuations, which can result in a significantly reduced neutrino mass sensitivity of KATRIN~\cite{Susanne}.

\subsection{The ECR method}
To mitigate the background problems arising from stored electrons in the KATRIN spectrometers, a novel method to remove stored electrons must be implemented. This method should be capable of removing electrons with energies up to several hundred keV, it should work with a very small duty cycle, and it should not enhance other eventual sources of background.

In the following we present a novel experimental method which has the promise to fulfill these requirements. It is based on stochastic heating of electrons by electron cyclotron resonance (ECR)~\cite{Holt, Geller}. The ECR technique is based on the principle that the frequency $f_{\text{RF}}$ of an external high frequency (RF) field is adjusted to the cyclotron frequency $f_{\text{c}}$ of stored electrons. In our specific case the resonance region is set to the center part of the spectrometer.  

In case the resonance condition $f_{\text{RF}} = f_{\text{c}}$ is met, the electron on average will gain a small amount of energy when passing the analyzing plane. Since a stored electron will pass the resonance zone $10^{4}$ times within a 10~ms period of stochastic heating, its energy will be significantly increased, growing by more than one order of magnitude. Consequently, the ECR technique  heats up electrons so that their cyclotron radius $r_{\text{c}}$ eventually becomes larger than the inner radius of the spectrometer walls ($r_{\text{s}}~=~5$~m), at which point they are absorbed.

In the Mainz neutrino mass experiment it was shown that background from stored electrons could be reduced by applying high frequency pulses to an electrode of the spectrometer~\cite{Mainz1,WeinheimerECR}. In this case, however, the electric field was mainly longitudinal to  the magnetic field, and the frequency ($1-2$~MHz) was much smaller than the electron cyclotron frequency corresponding to the magnetic field at the electrode position, so that this background reduction method is significantly different from the ECR method presented in our paper.

The paper is organized as following: first, we outline the basic principle of the ECR technique at a KATRIN spectrometer (section~\ref{single}). Then we discuss the experimental results obtained from test measurements of the ECR technique at the \prespectrometer{} (section~\ref{total}), demonstrating its highly effective background reduction capabilities. Finally, we use extensive MC simulations to derive optimum operating parameters for the ECR method at the main spectrometer.

\section{The working principle of ECR at a KATRIN spectrometer}\label{single}
The cyclotron frequency $f_{\text{c}}(B,\gamma)$ of a relativistic electron with Lorentz factor $\gamma$ is given by
\[
  f_{\text{c}}(B,\gamma) = \frac{\text{e} B}{2 \pi \text{m} \gamma},
\]
where $B$ denotes the external magnetic guiding field, and e and m represent the absolute value of the electron charge and its mass. 
Sources of primary stored electrons include $\beta$-decays of tritium molecules (HT, $\text{T}_2$) and processes occurring during $\alpha$-decays of the radon isotopes ${}^{219,220}$Rn. The energy distribution of these primary electrons varies by more than five orders of magnitude from the eV level up to about one hundred keV~\cite{Rn220ChargeDist, ConversionDataRn220, ConversionDataRn219, ShakeOff, KShakeOffRadon, ShellReorganization, LMShakeOffRadon}. However, in terms of cyclotron frequency $f_{\text{c}}(B,\gamma)$ this implies a variation of about 20\% ($1\leq\gamma\leq1.2$) only. The narrow frequency interval thus greatly facilitates the implementation of the RF-technique for removing electrons of very different energies. 

In the KATRIN spectrometers the magnetic field changes from its minimum value $B_{\text{c}}$ at the center to the ends of the vessel $B_{\text{e}}$ by four orders of magnitude (see figure~\ref{fig:KATRIN}). Consequently, the $f_{\text{c}}(B)$ dependence easily compensates the small $f_{\text{c}}(\gamma)$ variations, i.e.{} electrons of different energies will pass the resonance regions at different positions of the spectrometer. On the other hand, a magnetic field with steep gradients reduces the areas and times of resonance. Accordingly, we will show that by sweeping the RF-frequency the efficiency for stochastic heating of stored electrons can be improved significantly (see section~\ref{ssc:frequency}).

The net energy gain of an electron depends on the relative phase between the electron's cyclotron motion and the external RF-field. In the case of an RF-field with constant frequency applied in a constant magnetic field, no net energy gain can be achieved, due to a periodically shifting phase relation. At KATRIN, however, the phase between the electron cyclotron motion and the RF-field changes randomly: at each transition through the resonance-region the energy of the electron changes. As a consequence, the trapped electron will penetrate more or less deep into the high field regions, which in turn leads to a random phase change between the electron and the RF-field. 

In the KATRIN main spectrometer field layout, electrons with energies above about $E_{\text{max}}\approx100$~keV cannot be trapped, as either their cyclotron radius at the analyzing plane becomes too large, or their storage is no longer stable due to non-adiabatic effects~\cite{Susanne}. The stochastic heating of electrons thus has only to provide the energy difference between the starting energy and $E_{\text{max}}$.

Since the phase space for gaining energy is always larger than the constrained phase space for losing energy (the electron cannot have less than zero kinetic energy), a net energy gain is achieved after many resonance-passes. The basic principle of the ECR technique is illustrated in figure~\ref{fig:trapped_radon}. It shows the trajectory of a 5~keV trapped electron for stable storage conditions and in the presence of an RF-field. The trajectories were simulated with the KATRIN simulation software for field calculation and particle tracking \textsc{Kassiopeia}~\cite{Kassiopeia, Kassiopeia2, Susanne}.


\section{Proof of principle at the KATRIN \prespectrometer{}}\label{total}
Based on the general considerations presented above, an experimental program was initiated to verify the background reduction capabilities of the ECR method at the KATRIN \prespectrometer{}. In addition to a proof-of-principle demonstration, the measurements should establish that the background reduction can be achieved by rather short RF-pulses and that no other background sources are enhanced due to the RF-feed-in. On a more technical level, in view of the rather complex inner electrode systems of the spectrometers, the measurements should also demonstrate that the coupling of an external RF-field does not affect the integrity of these crucial system components.

\subsection{Experimental setup}
In the final KATRIN setup the \prespectrometer{} affords the option to work as a pre-filter for $\beta$-electrons, thereby reducing their flux by six orders of magnitude. In this case the \prespectrometer{} acts as an electrostatic filter with a fixed retarding voltage set to a value of 300~V below the endpoint. At this operating point, the energy resolution of the \prespectrometer{} of $\Delta E = 70$~eV is sufficient to leave the interesting high-energy part of the $\beta$-spectrum unmodified. Before being integrated in the full KATRIN setup, the \prespectrometer{} was operated as a stand-alone facility with an extensive suite of test measurements~\cite{Fraenkle, Susanne, FlorianDoktor, SusanneDoktor, Prall, Nancy}. The major tasks of this program were the development of advanced technologies and experimental methods that later can be applied to the much larger main spectrometer. 
 
\begin{figure}
\centering
\includegraphics[width = 0.75\textwidth]{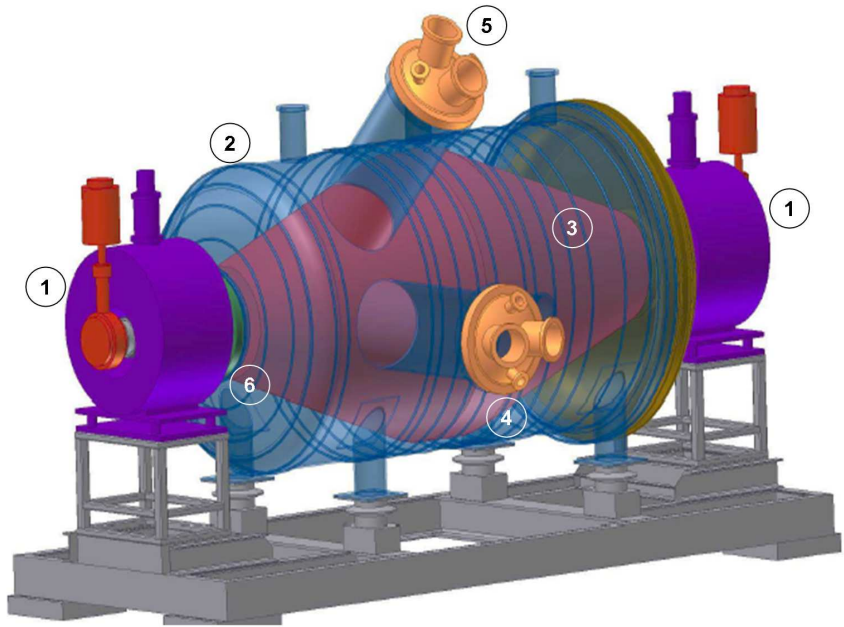}
\caption{Schematic view of the \prespectrometer{}. 1: superconducting solenoids. On the left end of the spectrometer an electron gun is installed to test the transmission properties~\cite{FlorianDoktor} (here the Pt30-2 krypton source is mounted); on the other end a 64-pixel Si-PIN diode detector is mounted, 2: \prespectrometer{} vessel, 3: inner electrode system, 4: $90^{\circ}$ pump port (housing the Au30-1 krypton source), 5: $45^{\circ}$ pump port, 6: insulator}
\label{fig:prespec}
\end{figure}

The \prespectrometer{} (see figure~\ref{fig:prespec}) has a length of 3.4~m and a diameter of 1.7~m. At both ends a superconducting magnet is installed providing a magnetic field of 4.5~T in the center of the solenoid and 15.6~mT in the center of the spectrometer. At one end an $8\times8$ Si-PIN diode array is mounted. As a novel design feature, when compared to the Mainz and Troitsk set-ups~\cite{Mainz1, Mainz2, Troitsk1, Troitsk2}, the vessel itself is set on (negative) high voltage. An inner electrode system consisting of two full electrodes and a wire electrode can be set to a different potential relative to the tank.

\begin{figure}
\begin{center}
\includegraphics[width = 0.5\textwidth]{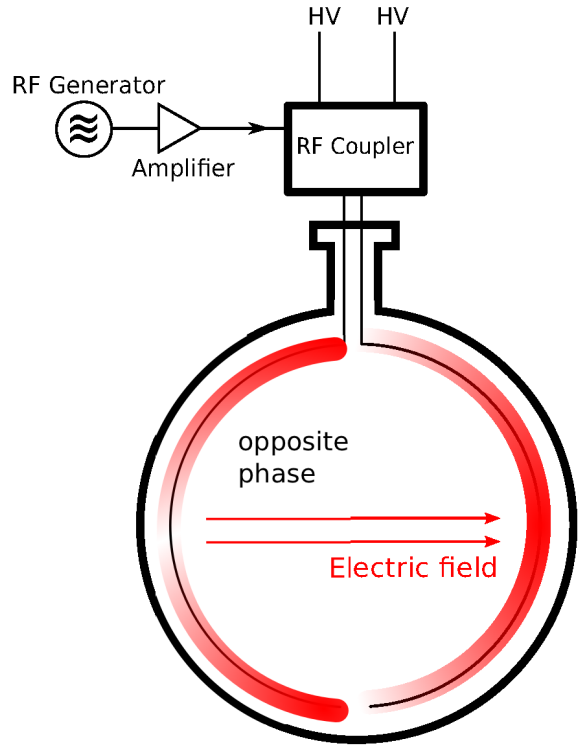}
\caption{Sketch of experimental setup: An RF-signal produced by an RF-generator is amplified and coupled into the inner electrode system of the \prespectrometer{}, which can be operated in a dipole mode, allowing the application of a transversal electric field.} 
\label{fig:setup}
\end{center}
\end{figure}

\begin{figure}
\centering
\begin{minipage}{0.52\textwidth}
  \includegraphics[width = \textwidth]{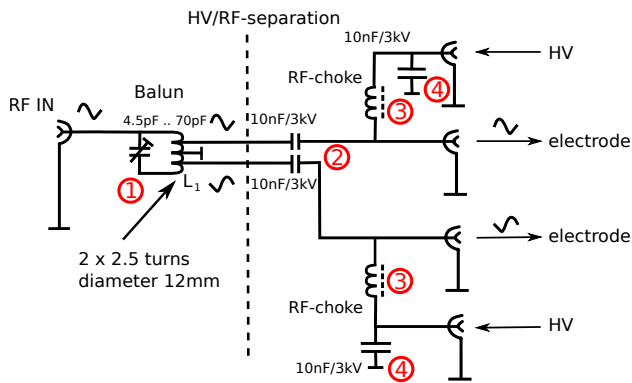}
\end{minipage}
\hfill
\begin{minipage}{0.47\textwidth}
  \includegraphics[width = \textwidth]{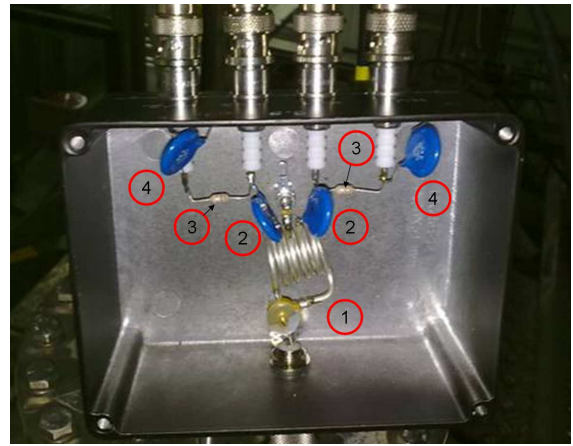}
\end{minipage}
\caption{Electric circuit diagram of the RF-coupler box, which allows for a superposition of HV and RF-components to supply the inner electrodes. Left: Schematic view. Right: Photograph of the coupler box.  1) The ``Balun'' unit changes the single-ended signal to a differential signal (two opposite phase signals), it is constructed as a resonant circuit as shown in the schematics. 2) Capacitors to separate the HV DC component from the RF-input. 3) RF-choke and 4) capacitors working as LC low pass filter to prevent the RF-component from entering the HV supplies.}
\label{fig:technicalimp}
\end{figure}

\begin{figure}
\begin{center}
\includegraphics[width = \textwidth]{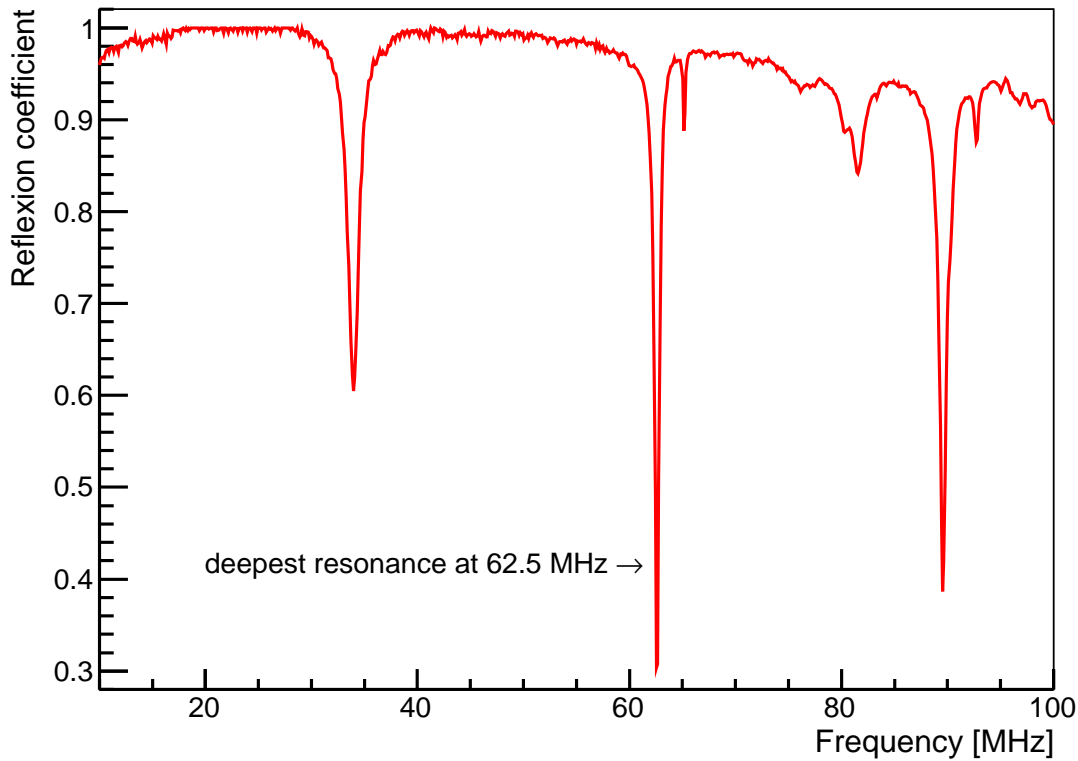}
\caption{The measured signal reflection coefficient as a function of frequency. A dip in the reflection coefficient marks an area of resonance. A sharp resonance is an evidence of a high magnification of the electric field inside the spectrometer. At a frequency of 62.5~MHz (corresponding to a wavelength of 4.8~m) 70\% of the applied energy was absorbed by the system.} 
\label{fig:reflectioncoefficient}
\end{center}
\end{figure}

The basic principle of the experimental setup for RF-injection at the \prespectrometer{} is shown in figure~\ref{fig:setup}. An RF-coupler box (see figure~\ref{fig:technicalimp}) superimposes the RF-signal from an RF-power amplifier (model: TP 30/10, manufacturer: Schlumberger) to the DC voltages required to shape the electrostatic potential. In order to obtain high RF-voltages on the inner electrodes, it was examined whether resonant amplification could be obtained by exploiting the natural electric resonance of the inner electrode system. Although this electrode system had not been designed ab initio with respect to RF-properties, the presence of symmetric dipole-electrodes turns out to be favorable. A network analyzer was used to examine the impedance behavior of the electrode system. Low-impedance resonant frequencies were found at 32~MHz, 62.5~MHz and 89~MHz (see figure~\ref{fig:reflectioncoefficient}). For the measurements reported on below, it was decided to use a fixed excitation frequency 
of $f_{\text{RF}}=62.5$~MHz. 

At this frequency, the dipole-electrodes act like a symmetric $\lambda$/2 transmission line (see figure~\ref{fig:setup}). The pair of electrode connection rods form, by coincidence, a  $\lambda$/4 transmission line feeding the $\lambda$/2 dipole from one end, transforming the high impedance at this point into a low impedance at the feed point within the RF-coupler box. The narrow resonance peak of $\sim2$~MHz width points to a high Q-factor, from which in turn a significant resonant amplification can be expected. The RF-power that was fed into the RF-coupler box was of the order of $6 - 8$~W ($17-20$~V RMS into 50~$\Omega$) only.

\begin{figure}
\begin{center}
\includegraphics[width = 0.8\textwidth]{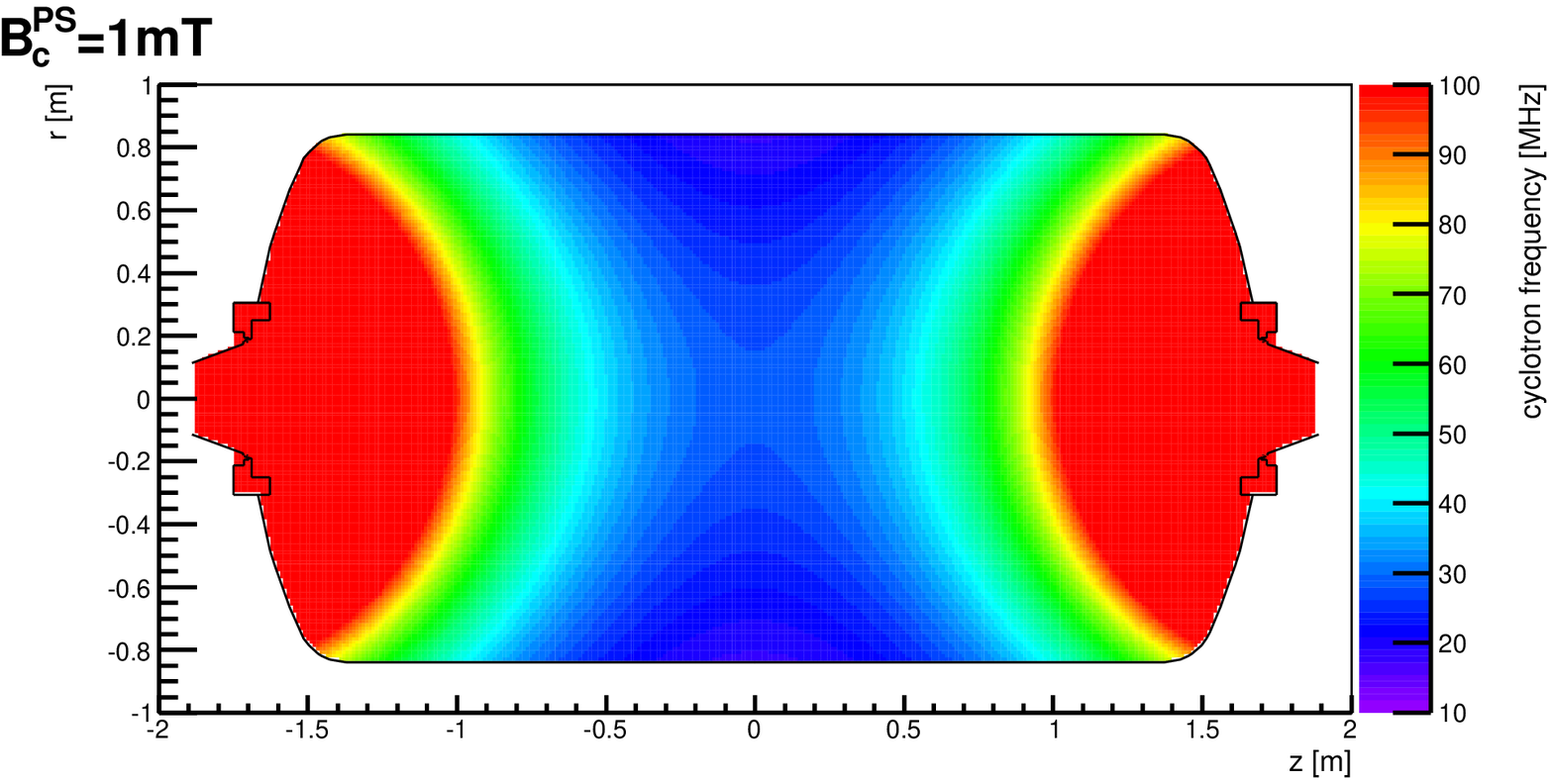}
\includegraphics[width = 0.8\textwidth]{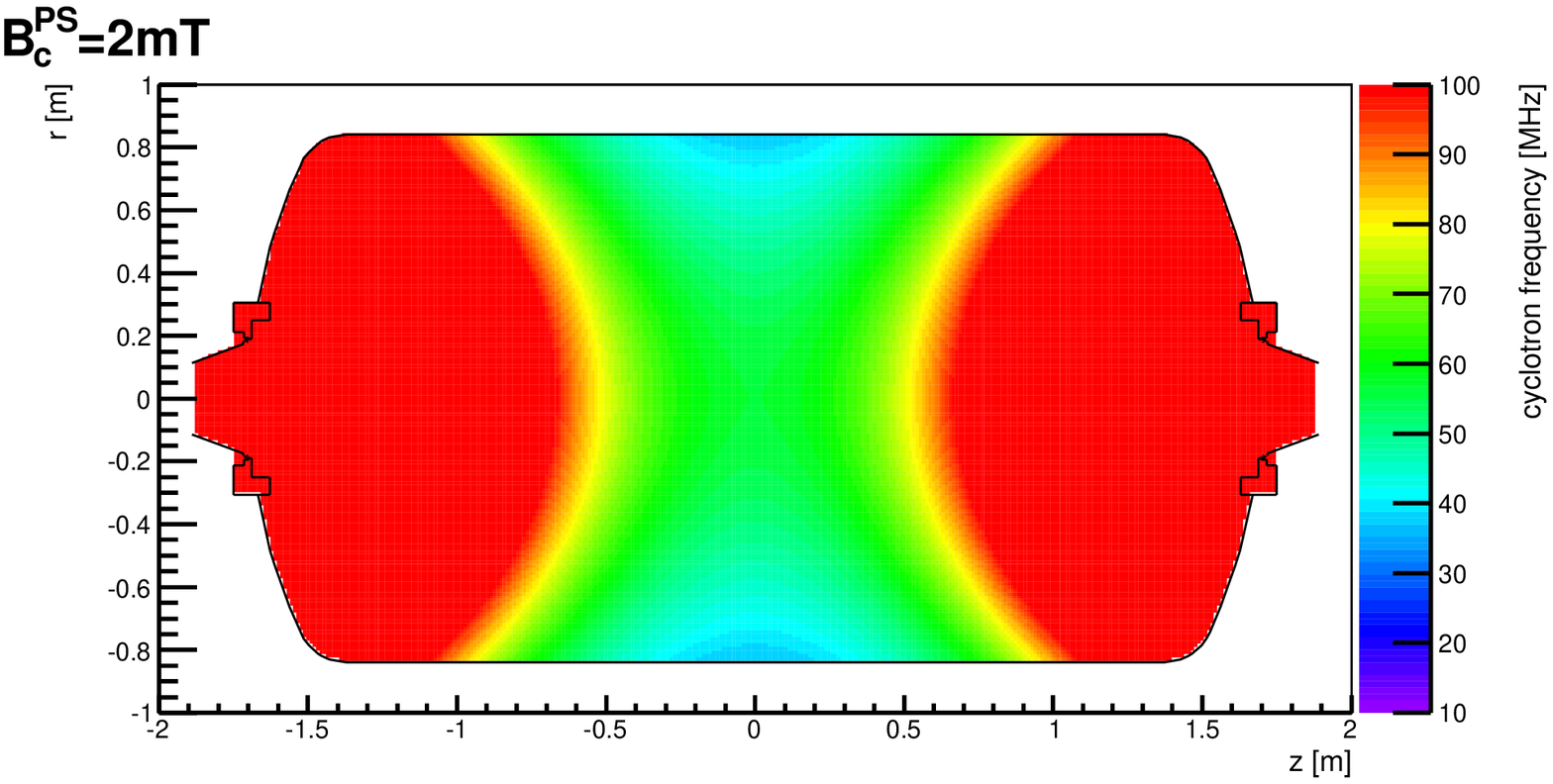}
\includegraphics[width = 0.8\textwidth]{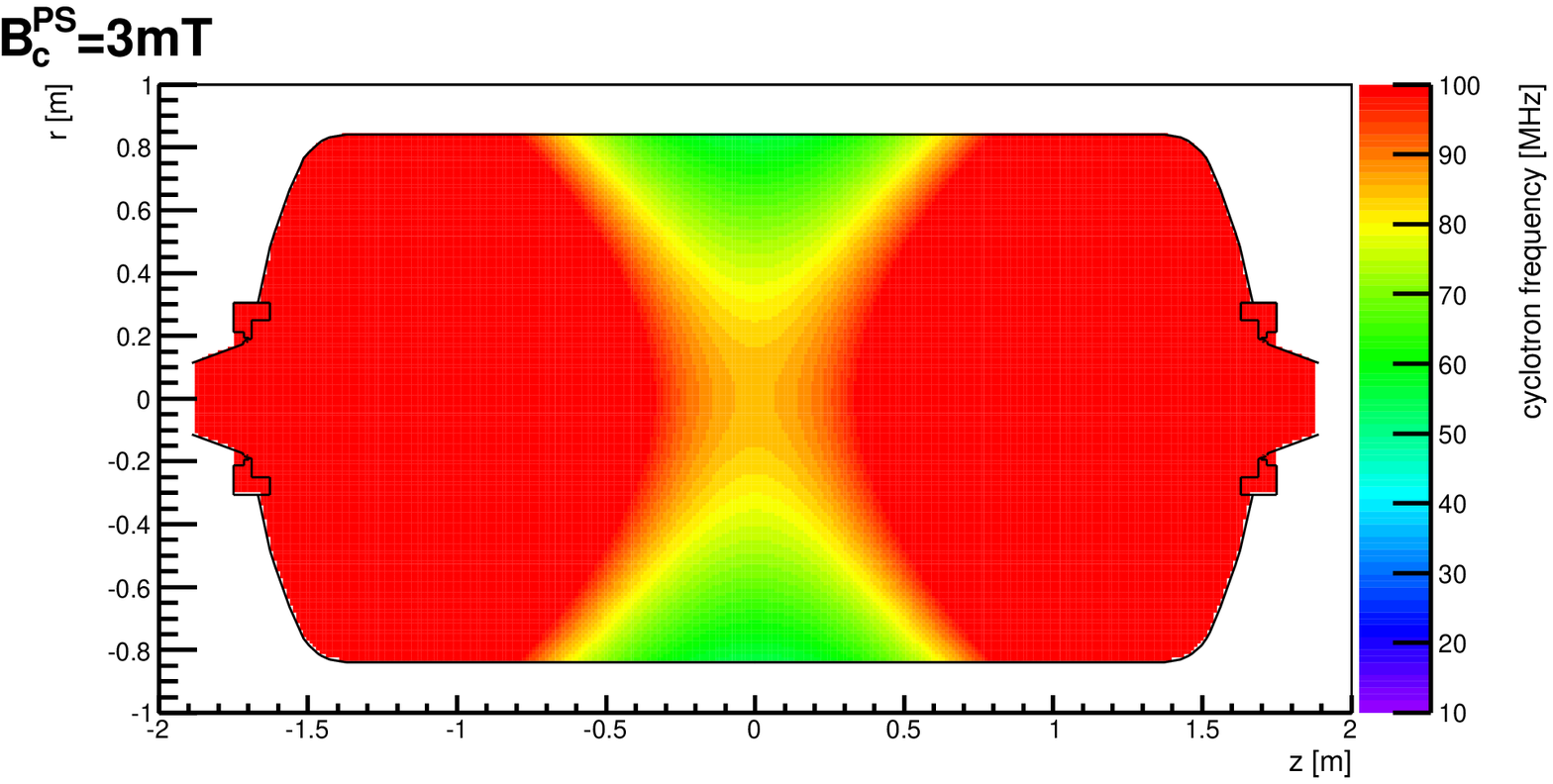}
\caption{Cyclotron frequency of a 1~keV electron in the \prespectrometer{} for three different values of $B_{\text{c}}^{\text{PS}}$.  In the green colored regions the cyclotron frequency $f_{\text{c}}$ coincides with the frequency of the external RF-field $f_{\text{RF}} = 62.5$~MHz. At a low magnetic field strength of $B_{\text{c}}^{\text{PS}}=1$~mT this only occurs in narrow regions close to the entrance and exit of the spectrometer. When increasing $B_{\text{c}}^{\text{PS}}$ to 2~mT, the area of resonance axially moves inward to the center and gets larger. Since electrons spend a large fraction of their axial oscillation period in the center and thus in resonance, a large effect is expected there. For magnetic field strengths of $B_{\text{c}}^{\text{PS}}=3$~mT and higher, the area of resonance gets smaller and moves radially outwards before finally disappearing.} 
\label{fig:resonancecondition}
\end{center}
\end{figure}

With the RF-frequency being fixed at $f_{\text{RF}}=62.5~$~MHz, the magnetic field at the center of the \prespectrometer{} $B_{\text{c}}^{\text{PS}}$ has to be adjusted to bring the electrons in resonance condition $f_{\text{c}}=f_{\text{RF}}$ with the external RF-field. Figure~\ref{fig:resonancecondition} shows the resonance conditions for different magnetic field configurations ($B_{\text{c}}^{\text{PS}} = 1, 2, 3$~mT). The calculation shows that the largest resonance area is expected for $B_{\text{c}}^{\text{PS}}=2.15$~mT. 

\subsection{Simulation software}
\label{ssc:simulationsoftware}
Accompanying the measurements, MC simulations were performed using the simulation software~\textsc{Kassiopeia}~\cite{Kassiopeia, Kassiopeia2}, which has been developed over the past years by the KATRIN collaboration. \textsc{Kassiopeia} is a software package that comprises precise and fast calculation of electromagnetic fields and particle trajectories. The trajectory calculations of \textsc{Kassiopeia} are based on explicit Runge-Kutta methods described in~\cite{RungeKutta, RungeKutta2, RungeKutta3}. Electric and magnetic field calculations are performed via the zonal harmonic expansion~\cite{FerencEl,FerencMag}. In the case of electric fields, computations are carried out using the boundary element method~\cite{BEM}. 

Elastic, electronic excitation and ionization collisions of electrons with molecular hydrogen are included in the \textsc{Kassiopeia}. Total and differential cross section for elastic scattering are taken from~\cite{scattering6,scattering7,scattering9}, data for inelastic scattering (electronic excitation and ionization) is based on~\cite{scattering1,scattering2,scattering3,scattering4,scattering5,scattering8}. Furthermore, arbitrary residual gas compositions of hydrogen, water, nitrogen or argon~\cite{CrossSectionHomePage,Argon} can be chosen. The field, tracking and scattering simulations originate from FORTRAN and C codes developed between 2000 and 2008 by one of us (F.{}~G.).

In the framework of the investigations presented below, \textsc{Kassiopeia} was equipped with a ${}^{83\text{m}}$Kr event generator taking into account the emission of conversion and Auger electrons, based on data in~\cite{ConversionDataKrypton, Penelope}. As a first approximation, the RF-field was implemented in \textsc{Kassiopeia} as a sinusoidal function of the form
\[
\label{RFField}
\vec{E}(x,t) = \vec{E}_{0}(x)\sin\left(\frac{f_{\text{RF}}(t) \cdot t}{2\pi}\right),
\]
with $\vec{E}_{0}(x)$ and $f_{\text{RF}}$ denoting amplitude and frequency of the RF-field, respectively. 
The amplitude $\vec{E}_{0}(x)$ is chosen to be oriented perpendicular to the beam axis ($E_0 = |\vec{E}_{0}|$), since only this component of the field is able to increase the transversal energy $E_{\perp}$ of the electrons. The frequency can either be stationary $f_{\text{RF}}(t) = f_{\text{RF}}^{\text{fix}}$, or swept through in steps $f_{\text{RF}}(t)=f_{\text{RF}}^{\text{sweep}}(t)$.

The functionality and validity of the implemented RF-field routines in \textsc{Kassiopeia} were tested by comparing the numerical results to analytical calculations based on~\cite{Geller} and~\cite{Holt}. In this context, we consider parameters like the energy gain, position and momentum of low-energy electrons in a constant magnetic field under the influence of an RF-electric field. The agreement between all parameters produced by \textsc{Kassiopeia} and the analytic solution is of the order of $10^{-10}$, which is of sufficient precision to validate the results presented in the following.  

\subsection{Source of stored electrons}
\begin{figure}
\centering
\begin{minipage}{0.49\textwidth}
  \includegraphics[width = 0.9\textwidth]{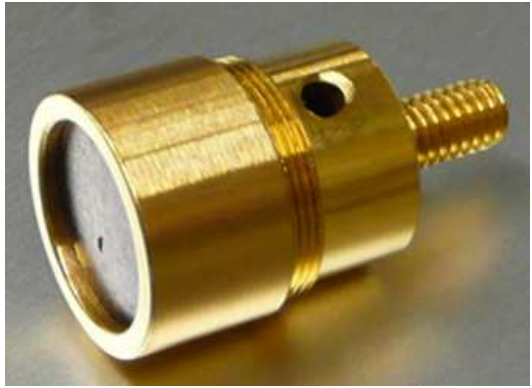}
\end{minipage}
\hfill
\begin{minipage}{0.49\textwidth}
  \includegraphics[width = \textwidth]{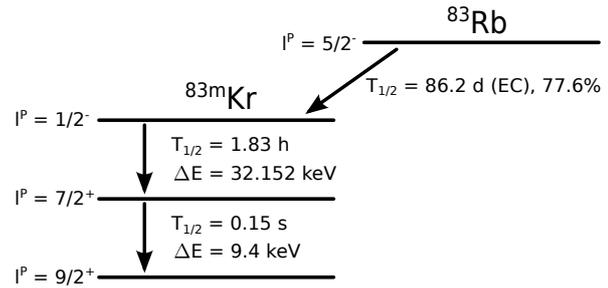}
\end{minipage}
\caption{Solid ${}^{83}$Rb/${}^{83\text{m}}$Kr source. Left: Photograph of the platinum-implanted ${}^{83}$Rb/${}^{83\text{m}}$Kr source. Right: Decay scheme of ${}^{83}$Rb.}
\label{fig:Krypton}
\end{figure}

To investigate the efficiency of the ECR technique in background reduction, an intense source of high-energy stored electrons was required. In this work a gold-implanted (Au30-1) and a platinum-implanted (Pt30-2) rubidium-krypton (${}^{83}$Rb/${}^{83\text{m}}$Kr) source were used (see figure~\ref{fig:Krypton})~\cite{Zboril}. The sources were installed at the horizontal pump port and at the electron gun position (see figure~\ref{fig:prespec}). The implanted mother isotope rubidium decays into an excited state of krypton (${}^{83\text{m}}$Kr), which is metastable and has a lifetime of 1.83~h (see figure~\ref{fig:Krypton}). The emanation rate of ${}^{83\text{m}}$Kr of the Au30-1 source is $c^{\text{Au}}_{\text{e}} = 11$\% and $c^{\text{Pt}}_{\text{e}} = 6$\% for the Pt30-2 source~\cite{ZborilDoktor}. 

The decay of ${}^{83\text{m}}$Kr produces conversion electrons in the multi-keV range (see decay scheme in figure~\ref{fig:Krypton}), leaving a vacancy in the electron shell. Subsequent relaxation processes lead to the emission of Auger electrons in an energy range of $10^2 - 10^4$~eV. As mentioned above, the creation of conversion electrons as well as full subsequent relaxation processes involving the emission of several Auger electrons in one decay are included in \textsc{Kassiopeia}. The average multiplicity per decay is $11.6\pm2.5$ electrons in our model.

\begin{figure}
\centering
  \includegraphics[width = 0.75\textwidth]{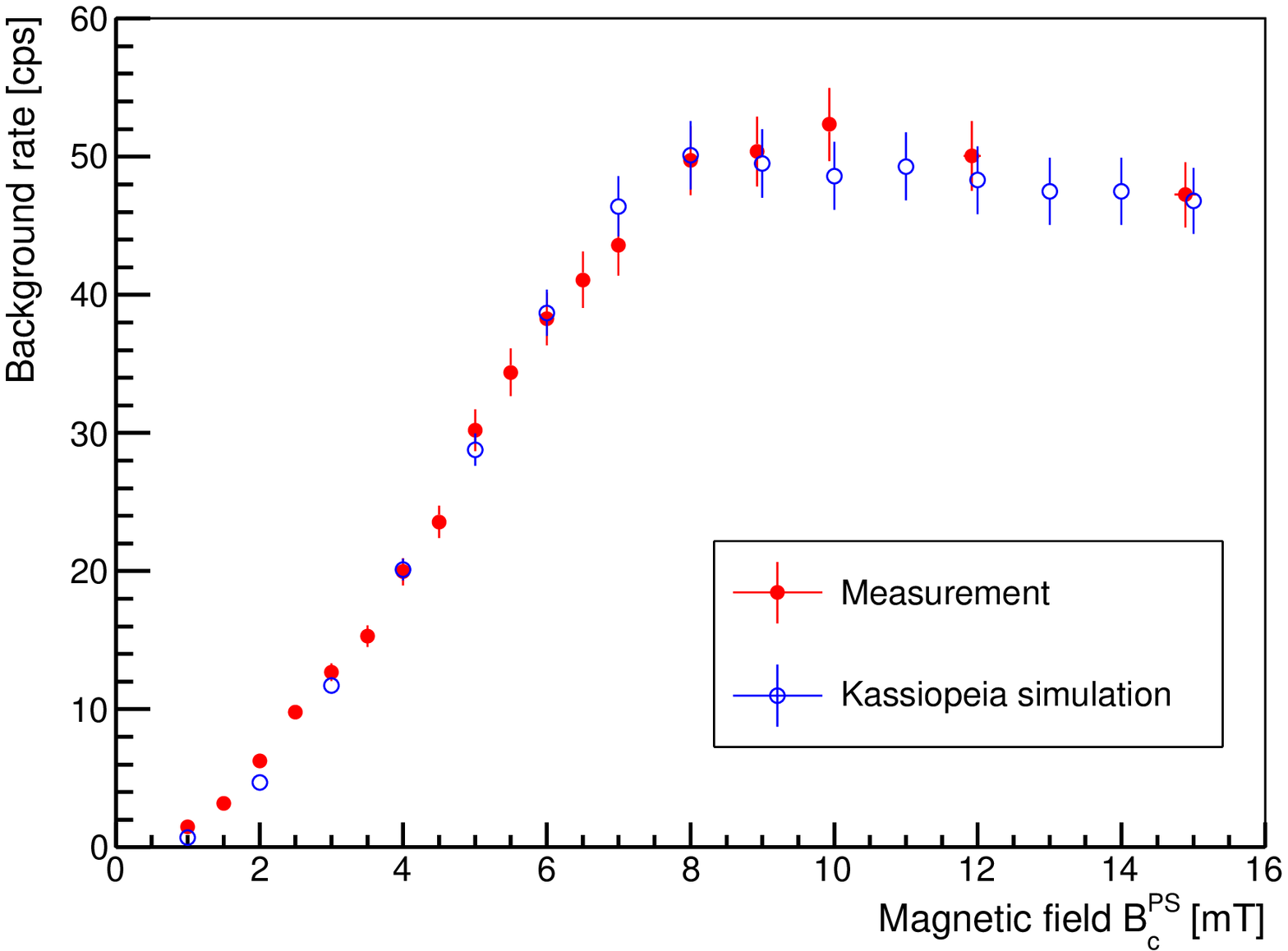}
  \includegraphics[width = 0.75\textwidth]{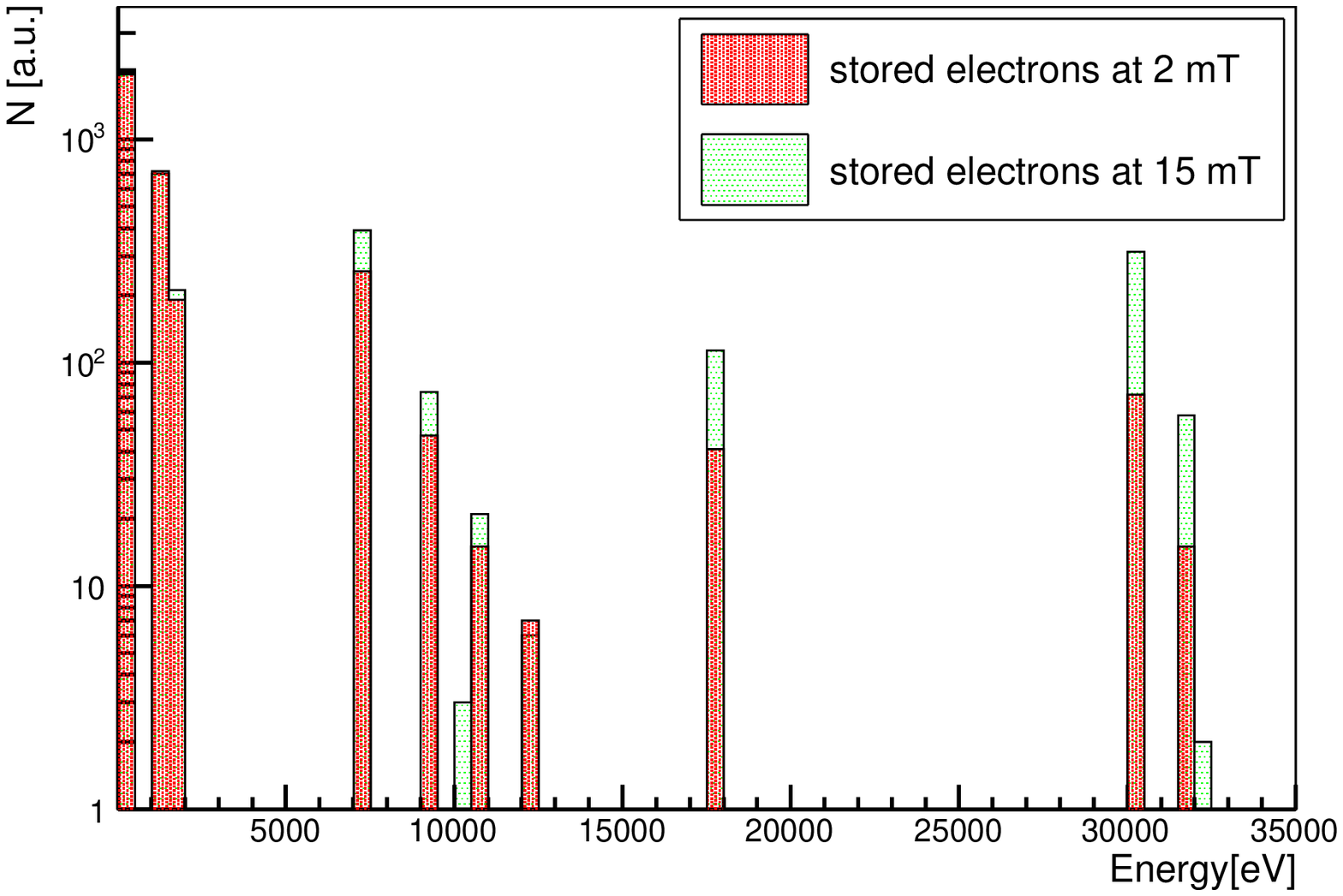}
\caption{Top: Measurement (full dots) and simulation (open dots) of the background rate for different central magnetic field strengths $B_{\text{c}}^{\text{PS}}$. The drastic drop of the rate with decreasing magnetic field is due to non-adiabatic effects. They occur when the magnetic field changes significantly within one gyration. In this case, the transformation of $E_{\perp}$ into $E_{\parallel}$ and vice versa is no longer proportional to the change of the magnetic field. Consequently, the electron's polar angle changes randomly and eventually hits a value below the trapping threshold. Bottom: Energy spectrum of stored electrons at $B_{\text{c}}^{\text{PS}} = 15$~mT compared to $B_{\text{c}}^{\text{PS}} = 2$~mT. From this plot it is evident that high energy electrons are affected most strongly by non-adiabatic effects: At low magnetic field, the number of high energy stored electrons is reduced by a factor of 10, whereas all low energy electrons are still stored.}
\label{fig:KryptonSpectrum}
\end{figure}

The number of ${}^{83\text{m}}$Kr decays per second $A_{\text{Kr}}$ of both sources in the sensitive spectrometer volume was determined by a comparison of the measurement of the background rate at different central magnetic fields $B_{\text{c}}^{\text{PS}}$ with a corresponding simulation (see figure~\ref{fig:KryptonSpectrum}). In the latter, the number of electrons reaching the detector per ${}^{83\text{m}}$Kr decay was simulated for different values of $B_{\text{c}}^{\text{PS}}$. Since $A_{\text{Kr}}$ is independent of $B_{\text{c}}^{\text{PS}}$, the simulated relative rate was fitted to the measured rate with $A_{\text{Kr}}$ as a free parameter. With a $\chi^2$ minimization the best fit value was found at $A_{\text{Kr}} = 0.13 \pm 0.002~\text{s}^{-1}$.

The resulting background level of $b_{\text{Kr}} = 6$~cps at $B_{\text{c}}^{\text{PS}} = 2.15$~mT is significantly larger than the normal background rate of $b_{\text{intr}} \approx 10^{-2}$~cps without the krypton source. The ECR technique is thus targeted at removing these stored electrons by stochastically heating them up. 

\subsection{Measurement results}
In the following we describe the experimental results obtained at the \prespectrometer{} operated in a mode where an external RF-field is applied to the inner electrode system. The objective of the first measurement was to verify a strong reduction of the background rate when adjusting the magnetic field $B_{\text{c}}^{\text{PS}}$ to the resonance condition $f_{\text{c}}=f_{\text{RF}}$. The second measurement was targeted at showing that the ECR technique is indeed removing the primary stored electrons. The final goal was to prove that the ECR technique does not amplify other potential sources of background.   

\subsubsection{Resonance effect}
The objective of this measurement was to demonstrate a resonance behavior of the background reduction R centered at $f_{\text{c}}=f_{\text{RF}}$, with R defined as 
\begin{equation}
 R = \frac{R_\text{RF, off}-R_\text{RF, on}}{R_\text{RF, on}}, 
\end{equation}
where $R_\text{RF, off}$ denotes the background rate when no RF-field is applied and $R_\text{RF, on}$ represents the rate with the RF-field applied. To find the maximal R, the magnetic field $B_{\text{c}}^{\text{PS}}$ and thereby the cyclotron frequency $f_{\text{c}}$ were increased from a small value up to the expected resonance at $B_{\text{res}}=2.15$~mT and above. 

\begin{figure}
\centering
\includegraphics[width = \textwidth]{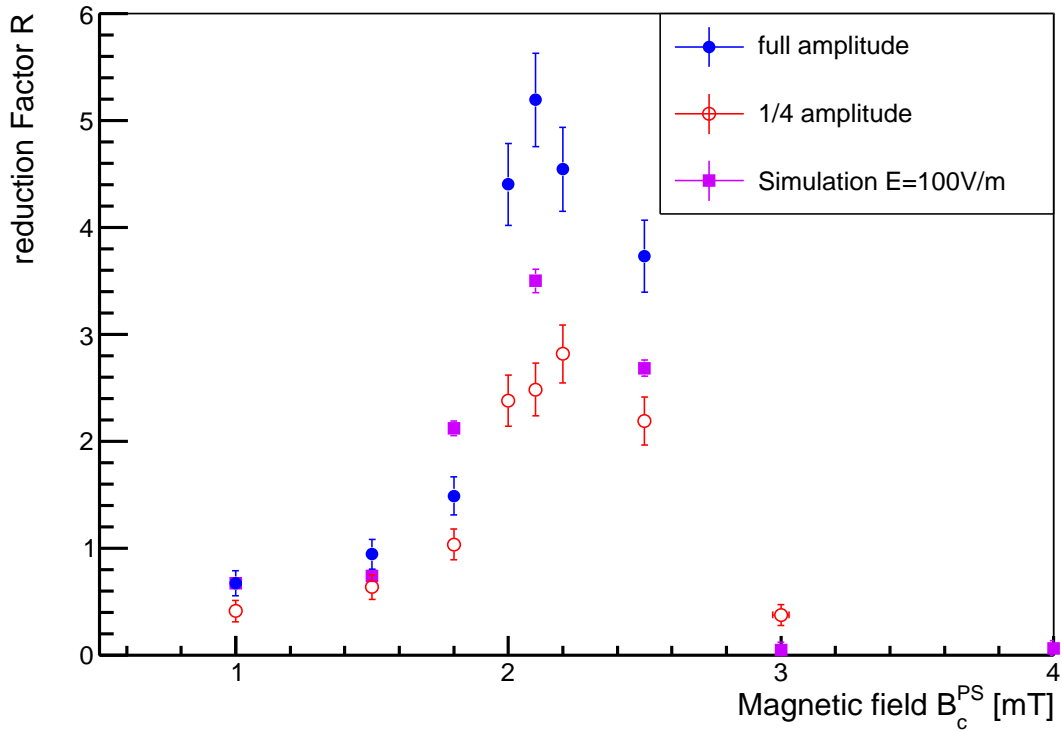}
\caption{Measurement of the relative reduction factor $R$ of the background from stored electrons as a function of the magnetic field strength in the center of the \prespectrometer{} $B_{\text{c}}^{\text{PS}}$. As expected, for the resonance condition $f_{\text{c}}=f_{\text{RF}}$ the maximal reduction is observed at $B_{\text{c}}^{\text{PS}}=2.15$~mT. The two RF-field amplitudes $E_0$ were controlled by adjusting the power fed into the system. The actual field inside the spectrometer can be determined by comparison to the simulation. The data show the expected resonance pattern of the ECR technique.} 
\label{fig:resoncemeasurement}
\end{figure}

Figure \ref{fig:resoncemeasurement} shows the experimental results of the background reduction factors R. The data feature a clear resonance centered at $B_{\text{res}} = 2.15$~mT, as expected for ECR. The asymmetric shape of the resonance curve can be explained by the inhomogeneity of the magnetic field (see figure~\ref{fig:resonancecondition}). Its minimum value $B_{\text{c}}^{\text{PS}}$ is reached in the center of the spectrometer, while increasing towards both ends. Correspondingly, if the central magnetic field is too low for resonance ($B_{\text{c}}^{\text{PS}} < B_{\text{res}}$), smaller resonant regions will exist only further away from the center. If the magnetic field is increased, these regions will move inwards and get larger, becoming maximal at the resonance field. However, if the central magnetic field is too high above the resonance value in the center ($B_{\text{c}}^{\text{PS}} \gg B_{\text{res}}$), the resonance condition $f_{\text{c}}=f_{\text{RF}}$ is not met in the entire spectrometer 
volume. This explains the asymmetric shape of the curve above resonance. The finite width of the cut-off above $B_{\text{res}}$ results from the different energies of stored particles. The scan of $B_{\text{c}}^{\text{PS}}$ through the resonance was performed with two different RF-amplitudes $E_0$. In both cases an identical resonance behavior was observed (see figure~\ref{fig:resoncemeasurement}). 

The reduction factors as well as the characteristic resonance pattern for low values of $B_{\text{c}}^{\text{PS}}$ and the asymmetric shape are reproduced by corresponding MC simulations using \textsc{Kassiopeia}~\cite{Kassiopeia,Kassiopeia2}. These simulations were also used to deduce the amplitude $E_0$ of the RF-field in the \prespectrometer{}, which could not be determined directly. As visible in figure~\ref{fig:resoncemeasurement}, an amplitude of $E_0 = 100$~V/m in the simulation yields comparable experimental reduction factors. This agreement implies actual field values of $E_0 = 100 - 200$~V/m at full amplitude.

Both measurement and MC simulations reveal that the fixed frequency setting used at the \prespectrometer{} only removes part of the stored electrons (factors of $3-5$), at resonance condition. The reason that not all stored electrons are removed is that we have used the technique of resonant amplification, which yields an RF-field of fixed frequency. Secondly, the rather steep gradients of the magnetic field of the \prespectrometer{} are not optimal for achieving large resonant areas for electrons over a broad energy range. At the main spectrometer the magnetic field is more homogeneous over larger areas, thus we expect the ECR technique to be much more efficient there.

\subsubsection{Effect of RF-pulsing}
To test whether the ECR technique is indeed removing the primary stored electrons, we investigated the behaviour of the background rate after RF-pulsing. If the primary electrons are removed, we expect that after an RF-pulse the background rate returns back to its original value following an inverse exponential behavior $R_{\text{off}} = 1-e^{-\frac{t}{t_{\text{r}}}}$. The parameter $t_{\text{r}}$ denotes the characteristic rise time of the rate, which is governed by the average storage time $t_{\text{stor}}$ of the trapped electrons and thus depends on the actual UHV conditions. 

\begin{figure}
\centering
\includegraphics[width = \textwidth]{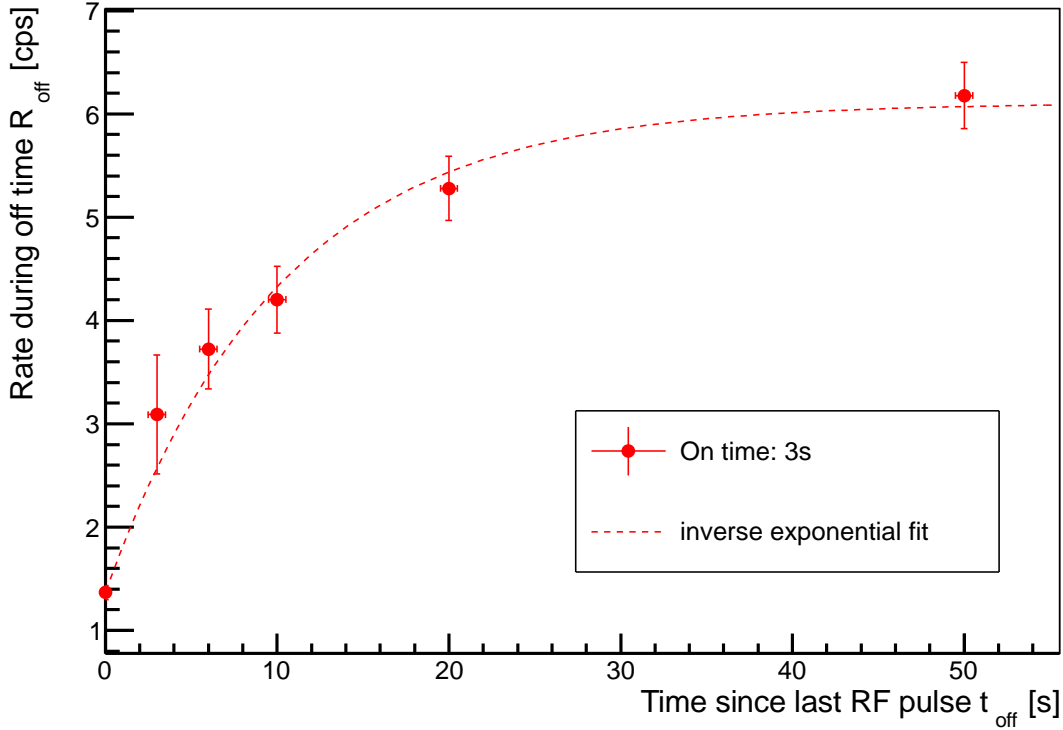}
\caption{Measurement of the increase of rate after 3~s long RF-pulses at a pressure of $p=(6\pm2)\cdot10^{-10}$~mbar and a magnetic field of $B_{\text{c}}^{\text{PS}}=2.15$~mT. Each data point represents a separate $\approx$ 10~h measurement during which the RF-pulse was alternatingly switched on for $t_{\text{on}} = 3$~s and switched off for $t_{\text{off}} = 5 - 50$~s, respectively. The plot shows the average rate during $t_{\text{off}}$. The rate $R_{\text{off}}(t_{\text{off}}=0\text{~s})$ of $1.37\pm0.07$~cps was measured with a continuous RF-field. The horizontal error bars indicate the inaccuracy of the timing, accordingly $R_{\text{off}}(t_{\text{off}} = 0\text{~s})$ has no horizontal error bar. An inverse exponential fit yields $t_{\text{r}} = (9.14\pm1.45)$~s. } 
\label{fig:cycling}
\end{figure}

Figure~\ref{fig:cycling} shows the measured behavior of $R_{\text{off}}$ as a function of the time $t_{\text{off}}$ since the last RF-pulse. An inverse exponential fit to the data yields $t_{\text{r}} = (9.14\pm1.45)$~s, which is in good agreement with the simulated average storage time of $t^{\text{sim}}_{\text{stor}} = (8.25\pm0.23)$~s. This value was achieved by a full MC simulation with \textsc{Kassiopeia} of 400 ${}^{83\text{m}}$Kr decays in the \prespectrometer{} at a pressure of $p=6\cdot10^{-10}$~mbar with a realistic gas composition of 50\% $\text{H}_2$, 25\% $\text{H}_2\text{O}$ and $\text{N}_2$ each. The parameter $t^{\text{sim}}_{\text{stor}}$ was deduced from an exponential fit to the storage time distribution of all stored electrons. The error only includes the statistical error from the fit, whereas systematic uncertainties such as uncertainties in the pressure or the energy distribution are not considered here.

As compared to the situation at the main spectrometer, typical storage times at the \prespectrometer{} measurements are much shorter. This is due to the higher pressure, and secondly because mostly low energy electrons are stored in the magnetic field configuration of the \prespectrometer{} with a central field of $B_{\text{c}}^{\text{PS}}=2.15$~mT (see figure~\ref{fig:KryptonSpectrum}). The cross section for ionization is maximal for $10^{2}$~eV electrons and decreases for higher energies~\cite{scattering5}. Therefore the cooling for low energy electrons is faster.

\subsubsection{Influence on other background sources}
Apart from stored electrons from nuclear decays studied above, an important background mechanism is related to cosmic muons~\cite{Benjamin, SusanneDoktor} that hit the spectrometer vessel and generate low-energy electrons. Starting from the inner surface, these electrons are electrically and magnetically shielded from the sensitive volume of KATRIN~\cite{Valerius}, since the inner electrode is on a more negative potential than the vessel hull and the magnetic field lines run parallel to the electrode surface. To test whether an RF-field influences this important shielding effect, ECR measurements without the krypton sources were performed at $B_{\text{c}}^{\text{PS}} = 2.15$~mT.

Without the krypton sources we measure an intrinsic background rate of $b_{\text{intr}} = (11.5 \pm 0.9)\cdot10^{-3}$~cps in the detector energy region of interest (ROI: 15~keV - 21~keV). The background rate was reduced to $b_{\text{intr}} = (6.2 \pm 1.2)\cdot10^{-3}$~cps when applying a permanent RF-field. Since the \prespectrometer{} background rate largely originates from the $\alpha$-decays of ${}^{219}$Rn, ${}^{220}$Rn emanating from the NEG strips and structural materials~\cite{Fraenkle, SusanneDoktor}, the reduction can likely be explained by the active removal of stored electrons following radon decays. 

From this result we can put a conservative upper limit on the cosmic muon induced background of $b_{\text{muon}} < (6.2 \pm 1.2)\cdot10^{-3}$~cps at the \prespectrometer{}, and we can exclude that this background is substantially increased by the application of the RF-field. This is an important result when discussing the use of the ECR method during the long-term scanning of the tritium $\beta$-spectrum with the main spectrometer.

\section{Expected background reduction at the main spectrometer}\label{consequences}
To validate and study in detail the concept of stochastic heating by the ECR method as a means of background reduction at the main spectrometer, extensive MC simulations were performed. These investigations are motivated by the expected background rates from stored electrons obtained by extrapolating the experimental results observed at the \prespectrometer{}~\cite{Fraenkle} to the significantly larger main spectrometer (diameter = 10~m, length = 24~m)~\cite{Susanne}. In the following we present the results of MC simulations with \textsc{Kassiopeia} that probe the efficiency of the ECR technique in alleviating this background at the main spectrometer.  

\textsc{Kassiopeia} was used to implement the full electromagnetic design of the main spectrometer, consisting of a system of superconducting coils and an external air coil system with a diameter of $d = 12$~m, which together provide an adiabatic guidance system for signal $\beta$-decay electrons through the spectrometers. The precision retarding potential is created in the following way: the spectrometer vessel itself is on negative high voltage (HV) and a wire-based inner electrode system is operated on a slightly more negative HV, allowing for a $10^{-6}$ precision of the filter potential~\cite{Valerius}. The RF-field was implemented as described in subsection~\ref{ssc:simulationsoftware}.

The aims of the simulations described below are to optimize the frequency settings, to test the efficiency of ECR with low field  amplitudes, to investigate the effects of ECR on electrons of different energies, and finally to analyze the impact of ECR on the statistical uncertainties in the neutrino mass analysis of KATRIN.

\subsection{Optimizing the frequency settings} \label{ssc:frequency}
To compare the two possible frequency settings ($f_{\text{RF}}^{\text{fix}}$ and $f_{\text{RF}}^{\text{sweep}}$) as well as to optimize the sweeping strategy we first carried out simulations with a simplified configuration in a constant magnetic field $B_{\text{const}}=0.3$~mT (matching the magnetic field in the center of the main spectrometer). The starting energy of the electrons was fixed to $E_{\text{prim}} = 5$~keV (matching the typical energies of electrons from nuclear decays). The cyclotron frequency of these electrons is $f_{\text{c}} = 6$~MHz, which corresponds to a cyclotron period of $t_{\text{c}} = 1.67 \cdot 10^{-7}$~s. 

\begin{figure}
\centering
\includegraphics[width=\textwidth]{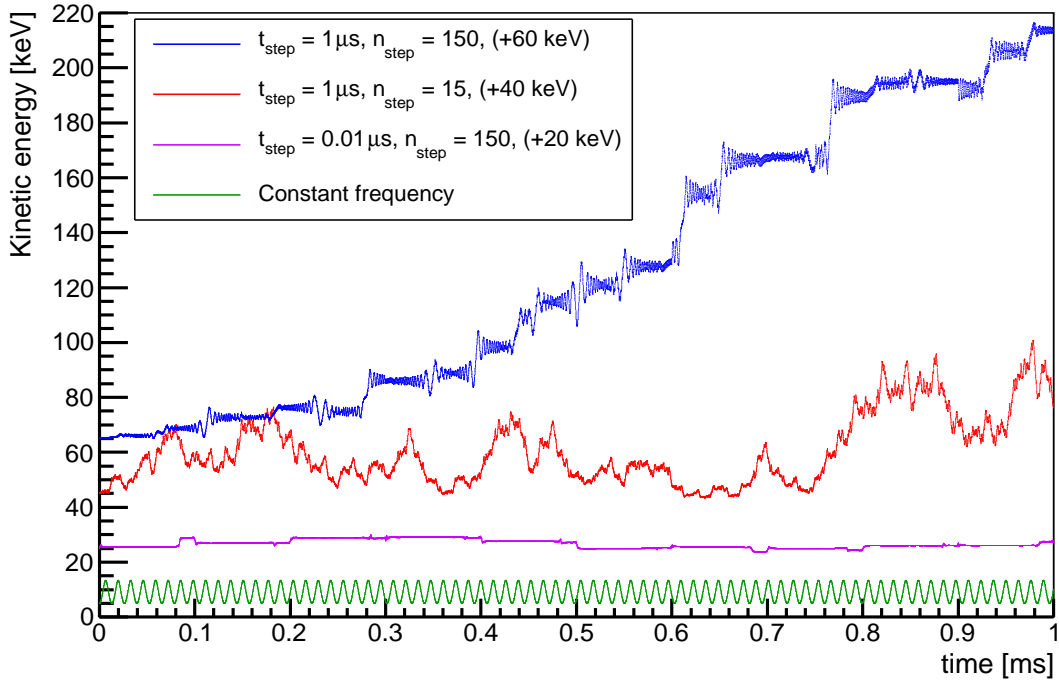}
\caption{Comparison of example frequency settings. The figure shows the kinetic energy of an electron with a starting energy $E_{\text{prim}} = 5$~keV in a constant magnetic field of $B_{\text{const}}=0.3$~mT in the presence of an RF-field with constant or sweeping frequency mode and an amplitude of $E_0 = 100$~V/m as a function of time. The constant frequency was set to $f_{\text{const}}=8.28$~MHz. This case (green) leads to a periodic energy change. In the sweeping mode the frequency was varied between $f_{\text{min}}=7$~MHz and $f_{\text{max}}=8.5$~MHz. The frequency is swept through in $n_{\text{step}}$ steps with an individual time step $t_{\text{step}}$. If the time steps are too short $t_{\text{step}} = 0.01~\mu$s (violet, +20~keV) or the number of sweeping steps is too small $n_{\text{step}} = 15$ (red, +40~keV), no significant energy increase can be achieved. The best result is obtained when the scanning of the frequency occurs in small steps with a duration of a few times the cyclotron period (blue, +60~keV). The energy curves for sweeping RF-cases have been separated by offsets of 20~keV, 40~keV and 60~keV, respectively, for better visibility.}
\label{fig:EvsT}
\end{figure}

As expected, the simulation demonstrates that no net energy gain can be achieved with a constant frequency RF-field in a constant magnetic field. Instead, the energy of the electron changes periodically with time (see figure~\ref{fig:EvsT}). In a constant magnetic field, a sweeping frequency mode is needed to heat up electrons. The frequency sweep is implemented as a stepwise increasing function, where the duration of a step $t_{\text{step}}$ and the number of steps $n_{\text{step}}$ within a sweep are free parameters. The simulations visualized in figure~\ref{fig:EvsT} show that the energy change of the electron by ECR is most efficient, if
\begin{itemize}
\item the duration of a single step is of the order of a few times the cyclotron period, i.e.{} $t_{\text{step}}\approx n\cdot t_{\text{c}}$.
\item $n_{\text{step}}$ is large so that many different frequencies occur within one sweep.
\end{itemize}

Following these initial results for a fixed energy and a constant magnetic field, we expand our investigations to the realistic case of the main spectrometer. The magnetic field there increases axially towards the field-generating solenoids. Consequently, the cyclotron frequency $f_{\text{c}}$ of trapped electrons changes along their trajectory. Therefore, an RF-field at a fixed frequency can lead to a net energy gain. By varying the frequency we expect to further improve the background reduction efficiency. This was investigated by considering the test case of electrons with energies up to 18.6~keV from tritium $\beta$-decay in the main spectrometer volume, both with constant and sweeping frequency of the RF-field.

The constant RF-frequency $f_{\text{RF}}^{\text{fix}}$ is chosen to correspond to the cyclotron frequency $f_{\text{c}}$ of a low-energy electron ($1$~eV) in the minimum magnetic field $B_{\text{c}}^{\text{MS}}$. In the sweeping frequency mode, the frequency $f_{\text{RF}}^{\text{sweep}}$ is changed between 8~and~10~MHz, corresponding to cyclotron frequencies of $1$~eV and $100$~keV electrons at $B_{\text{c}}^{\text{MS}}$, respectively. The parameters of the RF-field settings are summarized in table~\ref{tab:ECRSimSet}.

The comparison of the two RF-modes shows that in case of a sweeping frequency, all electrons could be removed, whereas in the case of constant frequency a fraction of 6.5\% of the electrons remains trapped within a fixed time interval of 500~ms (see table~\ref{tab:constsweep}). In case that not all electrons are removed within a single RF-period, the subsequent RF-pulses however will remove a significant fraction of these trapped electrons. The background level in between subsequent RF-pulses is thus unaffected by these electrons, which remain "parked" at high energies, as the energy-dependent cross section for ionizing collisions decreases in the keV-range~\cite{scattering5}.

\begin{table}
\caption{RF-field settings for the simulation with \textsc{Kassiopeia}}
\centering
\begin{tabular*}{\textwidth}{p{4.5cm}p{5cm}l}
\hline
\hline
RF-parameter & Frequency mode \\
\hline\hline
& constant & sweeping \\
\hline 
Frequency & $f=9.750$~MHz & $f_{\text{min}}=8$~MHz \\
		& & $f_{\text{max}}=10$~MHz \\
Time of step & - & $t_{\text{step}}=1\mu\text{s}$ \\
Number of steps & - & $n_{\text{step}}=200$ \\
Time of sweep & - & $ t_{\text{sweep}} = n_{\text{step}} \cdot t_{\text{step}} = 200 \mu\text{s}$\\	
RF-Amplitude & $E_0 = 100~\text{V/m}$ & $E_0 = 100~\text{V/m}$\\ 
\hline
\end{tabular*}
\label{tab:ECRSimSet}
\end{table}

\begin{table}
\caption{Percentage of electrons remaining trapped in the presence of an RF-field with different settings. The cases of no RF-fields and RF-fields with constant or sweeping frequency are compared. For these investigations an ensemble of $2\cdot10^3$~tritium $\beta$-decay electrons was simulated for each RF-configuration. Electrons are considered to be in stable storage condition if their storage time exceeds 500~ms. Without the RF-field, a fraction of 51.5\% of the $\beta$-electrons are stored, the remainders immediately hit the wall due to their large cyclotron radius, or leave the spectrometer axially due to non-adiabatic effects~\cite{Susanne}. The table demonstrates the high efficiency of the ECR technique, in particular in the sweeping mode with a complete reduction of stored electrons.}
\centering
\begin{tabular*}{\textwidth}{p{4.5cm}l}
\hline
\hline
RF-setting & Remaining percentage of trapped electrons \\
\hline\hline
off & 51.5\% \\
constant frequency & 6.5\% \\
sweeping frequency & 0\% \\ 
\hline
\end{tabular*}
\label{tab:constsweep}
\end{table}

\subsection{Optimizing the amplitude of the RF-field}
In order to comply with the long-term integrity of the main spectrometer wire electrode system with its 24000 wires of a diameter of 300~$\mu$m (outer layer) and 200~$\mu$m (inner layer)~\cite{Valerius}, the lowest possible amplitude of the RF-field $E_0$ is mandatory. For low $E_0$ values, the electrons have to pass the analyzing plane more often to reach the extraction energy of $E_{\text{max}}\approx 100$~keV and, consequently, longer removal times are expected for lower amplitudes.

\begin{figure}
\centering
\includegraphics[width = \textwidth]{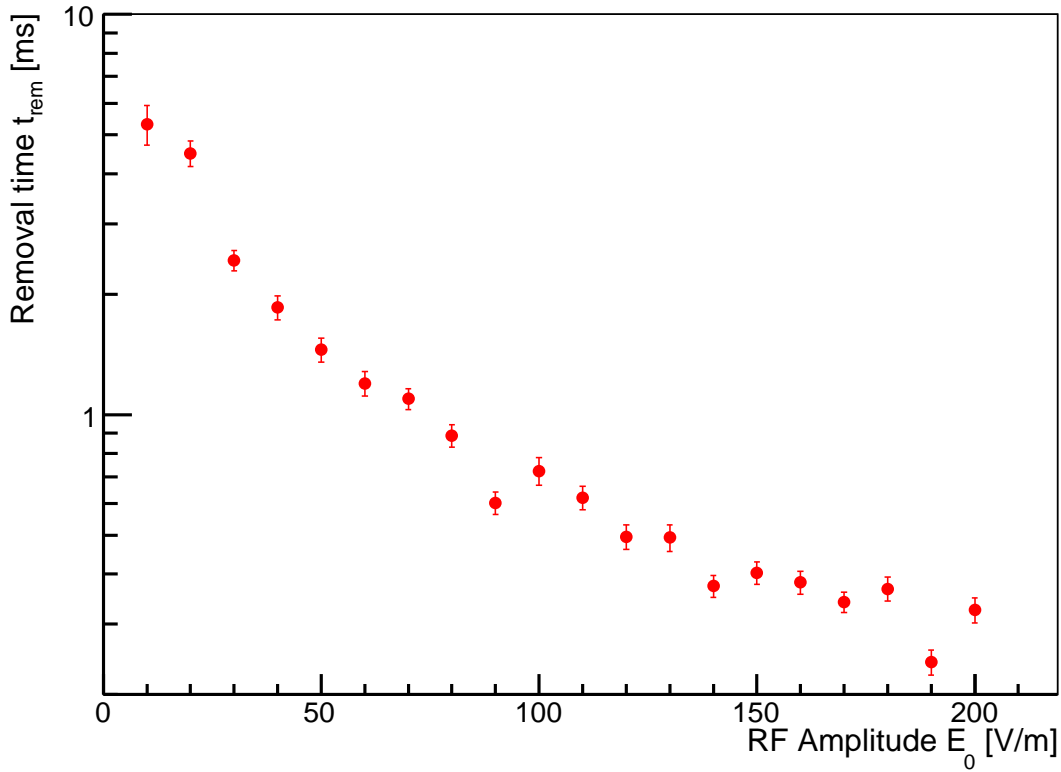}
\caption{Removal times $t_{\text{rem}}$ of tritium $\beta$-decay electrons in the presence of an RF-field with sweeping frequency as a function of RF-amplitude $E_0$. For this figure, $10^4$~tritium $\beta$-decay electrons were simulated for each $E_0$. The parameter $t_{\text{rem}}$ is given by the lifetime $\tau$, which was determined by fitting an exponential curve to the storage time distribution of the electrons for each $E_0$. The figure shows that $t_{\text{rem}}$ decreases for larger values of $E_0$. However, even for very low values of $E_0 = 10$~V/m the removal time $t_{\text{rem}}$ remains smaller than 10~ms.}
\label{fig:Run5_StorageTime}
\end{figure}

To test this hypothesis for trapped electrons from tritium $\beta$-decay, the effect of a sweeping RF-field with 20 different amplitudes between $E_0 = 10 - 200$~V/m was tested (the settings of the sweeping frequency are given in table~\ref{tab:ECRSimSet}). As an important result we find that for the values of $E_0$ considered all stored electrons could be removed. The average removal time increases by about an order of magnitude for smaller values of $E_0$, as expected (see figure \ref{fig:Run5_StorageTime}). However, even for very small amplitudes of $E_0 = 10$~V/m the removal time is still smaller than 10~ms.

\subsection{Effect of an RF-field on electrons of different energies}
During the long-term tritium scanning measurements the background of stored electrons will not only result from the above considered tritium $\beta$-decays in the spectrometer volume, but will also comprise electrons from  ${}^{219, 220}$Rn $\alpha$-decays with energies of up to 450~keV~\cite{ConversionDataRn220, ConversionDataRn219}. Consequently, we investigate the effectiveness of an RF-field as a function of the initial electron energy $E_{\text{prim}}$.

For this investigation, electrons of nine different starting energies in a range between $E_{\text{prim}} = 10~\text{eV} - 100$~keV were simulated in the presence of an RF-field with sweeping frequency (the parameters are given in table~\ref{tab:ECRSimSet}). For each of these values, an ensemble of $2\cdot10^4$ electrons was started isotropically in the main spectrometer. Again, the ECR method works efficiently, as even for this extended energy range all stored electrons were removed within 1~ms. This demonstrates that the ECR method is a promising tool to remove both electrons from tritium $\beta$-decays as well as from ${}^{219, 220}$Rn $\alpha$-decays. 

\subsection{Impact of ECR on statistical uncertainty of KATRIN}
In \cite{Susanne} it was demonstrated that the background due to stored electrons following ${}^{219,220}$Rn $\alpha$-decays and tritium $\beta$-decays results in a significant increase of the statistical uncertainty $\sigma^{\text{nucl}}_{\text{stat}}$ of the observable $\text{m}^2_{\overline{\nu}_e}$ of KATRIN. In the following we investigate how efficiently the application of the ECR technique can reduce $\sigma^{\text{nucl}}_{\text{stat}}$ and thereby improve the statistical neutrino mass sensitivity. 

\begin{figure}
\centering
\includegraphics[width = \textwidth]{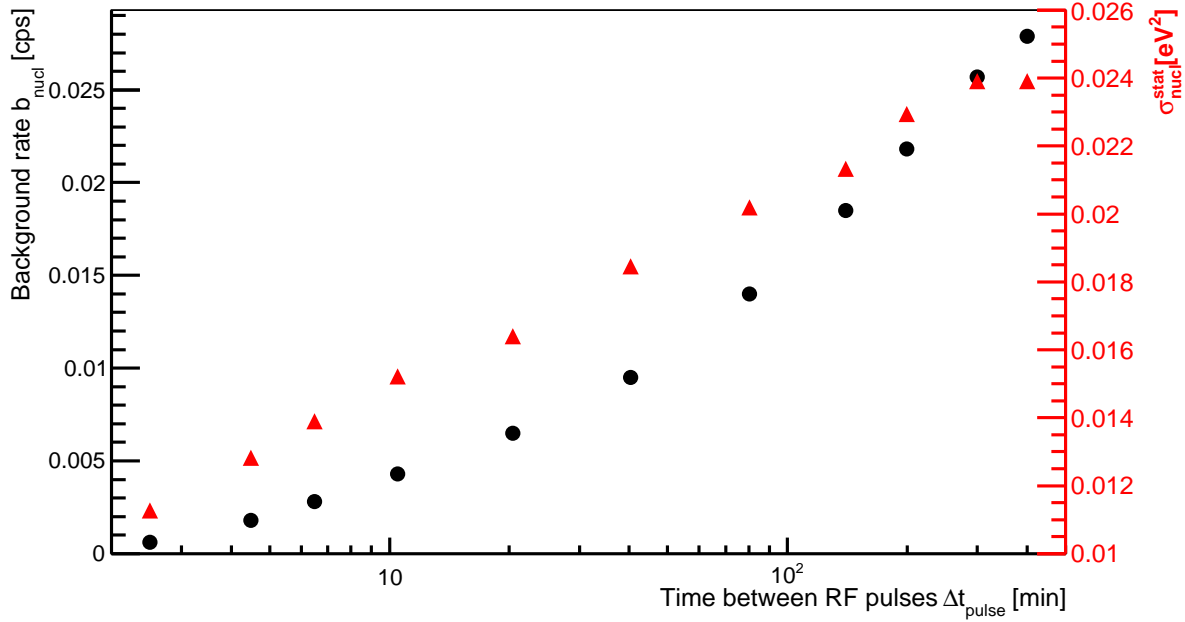}
\caption{The figure shows the background arising from nuclear decays $b_{\text{nucl}}$ (left axis, circles) and the associated statistical uncertainty $\sigma^{\text{nucl.}}_{\text{stat}}$ of the observable $\text{m}^2_{\overline{\nu}_e}$ (right axis, triangles) as a function of the time between RF-pulses $\Delta t_{\text{pulse}}$. A background model of $3\cdot10^{-2}$~cps without ECR is assumed. A value of $\Delta t_{\text{pulse}} = 2$~min is sufficient to reduce the background to less than $10^{-3}$~cps.}
\label{fig:ECRStat}
\end{figure}

As shown in the sections above, the ECR technique can be characterized by an almost negligible removal time $t_{\text{rem}} < 10$~ms and a 100\% efficiency of removing stored electrons of all energies. The only free parameter that influences the background level from nuclear decays is thus the time interval between two subsequent RF-pulses $\Delta t_{\text{pulse}}$. Since during the RF-pulsing no neutrino mass data can be taken, one aims to maximize $\Delta t_{\text{pulse}}$. Accordingly, we present the background level $b_{\text{nucl}}$ arising from nuclear decays and $\sigma^{\text{nucl}}_{\text{stat}}$ as a function of $\Delta t_{\text{pulse}}$ (see figure~\ref{fig:ECRStat}). 

Due to the good UHV conditions at the main spectrometer ($p = 10^{-11}$~mbar), the ionization times $t_{\text{ion}}$ of keV electrons are rather long (e.g. for a 5~keV electron $t_{\text{ion}} \approx 3$~min at $p = 10^{-11}$~mbar). We thus expect that correspondingly large values of $\Delta t_{\text{pulse}}$ are sufficient to reduce substantially the background rate. 

For the statistical analysis we use the detailed background model described in~\cite{Susanne}, which is based on MC simulations of ${}^{219,220}$Rn $\alpha$-decays and tritium $\beta$-decays in the main spectrometer and experimental data in~\cite{Fraenkle}. We expect a background rate from these decays of $b_{\text{nucl}} = 3\cdot10^{-2}$~cps~\cite{Susanne}. 

The statistical uncertainty $\sigma^{\text{nucl.}}_{\text{stat}}$ is determined by a $\chi^2$ fit of the theoretical integral $\beta$-spectrum for $5\cdot10^3$ simulated KATRIN measurements, considering the nuclear decay model as the only background source. For each simulation $\text{m}_{\overline{\nu}_e} = 0$~eV is assumed. The width of the distribution of the fitted neutrino mass squared $\text{m}^2_{\overline{\nu}_e}$ determines $\sigma^{\text{nucl.}}_{\text{stat}}$. The statistical analysis, visualized in figure~\ref{fig:ECRStat}, reveals following facts: 
\begin{itemize}
    \item With a value of $\Delta t_{\text{pulse}} = 2$~min the background from stored electrons can be eliminated almost completely to a level of $b_{\text{nucl}} = 6\cdot10^{-4}$~cps. The statistical uncertainty of $\sigma^{\text{nucl.}}_{\text{stat}}$ = 0.011 associated with this background is comparable to the situation without background where the statistical error is dominated by the variation of the signal count rate. With a pulse length of $t_{\text{p}} = 10$~ms and a repetition time of $\Delta t_{\text{pulse}} = 2$~min a duty cycle of $t_{\text{p}}/(\Delta t_{\text{pulse}} + t_{\text{p}})\approx8\cdot10^{-5}$ would be reached, which implies a negligible loss of neutrino mass measurement time. Moreover, this RF-pulsing rate ideally fits to the measurement schedule of tritium scanning~\cite{DesignReport} where the retarding potential is varied on the order of minutes. Accordingly, the RF-pulses could be applied while setting the retarding potential.
	\item Repetition times of $\Delta t_{\text{pulse}} = 10$~min reduce the original background level of $b_{\text{nucl}} = 3\cdot10^{-2}$~cps to $b_{\text{nucl}} = 4.3\cdot10^{-3}$~cps, corresponding to a statistical uncertainty of $\sigma^{\text{nucl}}_{\text{stat}}$ = 0.016. As a comparison, this uncertainty is of the same size as the statistical uncertainty associated with a constant (Poisson distributed) background level of $10^{-2}$~cps as anticipated in~\cite{DesignReport}.
	\item For long repetition times $\Delta t_{\text{pulse}} > 6$~h the ECR technique is no longer effective.
\end{itemize}
These considerations are also valid for all other active background reduction mechanisms that allow for complete removal of all trapped particles in time periods of the order of ms.

\section{Conclusion}
Previous measurements~\cite{Fraenkle} and simulations~\cite{Nancy, Susanne} have revealed that stored electrons arising from nuclear decays in the volume of the KATRIN spectrometers can cause large background rates that exceed the design background limit of $10^{-2}$~cps. It was shown that the non-Poissonian nature of this background can significantly influence the neutrino mass sensitivity of KATRIN.

In this paper, we have demonstrated the feasibility of a novel background reduction technique based on stochastic heating of stored electrons via the process of electron cyclotron resonance (ECR). 

Measurements at the KATRIN \prespectrometer{} have proven the basic functionality of the ECR method. In particular, we have demonstrated that rather short RF-pulses of $\Delta t\leq2$~s are sufficient to achieve a large background reduction factor of about five. Moreover, we could show that the RF-field does not increase backgrounds from other sources, in particular muon-induced background. 

In addition to these measurements, we have carried out extensive simulations of the ECR technique with the \textsc{Kassiopeia} code to study the efficiency of background reduction at the main spectrometer. Two RF-operation modes were considered: a mode with constant frequency, as well as a mode with sweeping frequency. The simulations revealed the sweeping frequency mode yields a complete background reduction, while the technologically simpler constant frequency mode still removes a large part (93.5\%) of stored electrons. Moreover, the simulations with \textsc{Kassiopeia} proved that the stochastic heating is efficient for electrons with energies up to 100~keV, thus covering the entire energy range of stored electrons from tritium $\beta$-decays and ${}^{219, 220}$Rn $\alpha$-decays.

Most importantly, the simulations showed that even in the case of a very low RF-amplitude of only $E_0=10$~V/m, which is preferable with respect to the integrity of the wire electrode, all stored electrons can be removed within less than 10~ms. Furthermore, statistical simulations revealed that the time between two subsequent RF-pulses of $\Delta t_{\text{pulse}} \approx 2$~min suffices to almost completely eliminate the contribution from the background due to stored electrons to the statistical error budget of KATRIN. The low RF-amplitudes and the small duty cycle of $\approx8\cdot10^{-5}$ are important prerequisites to assure the integrity of the wire electrode system while at the same time not affecting the measuring time of KATRIN.

Based on the experimental work at the \prespectrometer{} presented here, as well as on the corresponding MC simulations described above, we intend to test and optimize the ECR method, as well as other active reduction methods at the upcoming commissioning measurements at the KATRIN main spectrometer, with the aim of operating the spectrometer nearly free of background.

\section*{Acknowledgements}
This work has been supported by the Bundesministerium f{\"u}r Bildung und Forschung (BMBF) with project number 05A08VK2 and the Deutsche Forschungsgemeinschaft (DFG) via transregio 27 ``Neutrinos and beyond''. We also would like to thank Karlsruhe House of Young Scientists (KHYS) of KIT for their support (S.M., D.F., N.W.). For his technical support we would like to thank H. Skacel.

\section*{References}

\bibliography{references}

\begin{thebibliography}{10}

\bibitem{DesignReport}
J.~Angrik et~al.
\newblock {KATRIN Design Report}.
\newblock FZKA-7090, {2004}.

\bibitem{OttenWeinheimer}
E.~W. Otten and C.~Weinheimer.
\newblock {Neutrino mass limit from tritium $\beta$-decay}.
\newblock {\em Reports on Progress in Physics}, 71:086201, 2008.

\bibitem{WGTS}
W.~K{\"a}fer.
\newblock {The Windowless Gaseous Tritium Source of KATRIN}.
\newblock {\em Progress in Particle and Nuclear Physics}, 64(2):297--299, 2010.
\newblock Neutrinos in Cosmology, in Astro, Particle and Nuclear Physics,
  International Workshop on Nuclear Physics, 31st course.

\bibitem{MACE1}
V.~M. Lobashev and P.~E. Spivak.
\newblock A method for measuring the electron antineutrino rest mass.
\newblock {\em Nuclear Instruments and Methods in Physics Research Section A:
  Accelerators, Spectrometers, Detectors and Associated Equipment},
  240(2):305--310, 1985.

\bibitem{MACE2}
A.~Picard et~al.
\newblock {A solenoid retarding spectrometer with high resolution and
  transmission for keV electrons}.
\newblock {\em Nuclear Instruments and Methods in Physics Research Section B:
  Beam Interactions with Materials and Atoms}, 63(3):345--358, 1992.

\bibitem{MagneticMirror1}
H.~Higaki, K.~Ito, K.~Kira, and H.~Okamoto.
\newblock {Electrons Confined with an Axially Symmetric Magnetic Mirror Field}.
\newblock {\em AIP Conference Proceedings}, 1037(1):106--114, 2008.

\bibitem{MagneticMirror2}
T.~Tsuboi, E.~Y. Xu, Y.~K. Bae, and K.~T. Gillen.
\newblock {Magnetic bottle electron spectrometer using permanent magnets}.
\newblock {\em Review of Scientific Instruments}, 59(8):1357--1362, 1988.

\bibitem{MagneticMirrorBook}
U.~S. Inan and M.~Golkowski.
\newblock {\em {Principles of Plasma Physics for Engineers and Scientists}}.
\newblock Cambridge, 2011.

\bibitem{scattering1}
J.~W. Liu.
\newblock {Total cross sections for high-energy electron scattering by
  $\mathrm{H}_{2}$ (${}^{1}\Sigma_{g}^{+}$), $\mathrm{N}_{2}$
  (${}^{1}\Sigma_{g}^{+}$), and $\mathrm{O}_{2}$
  (${}^{3}\Sigma_{g}^{\mathrm{-}}$)}.
\newblock {\em Phys. Rev. A}, 35:591--597, Jan 1987.

\bibitem{scattering2}
J.~W. Liu.
\newblock {Total Inelastic Cross Section for Collisions of $\mathrm{H}_{2}$
  with Fast Charged Particles}.
\newblock {\em Phys. Rev. A}, 7:103--109, Jan 1973.

\bibitem{scattering3}
W.~Hwang et~al.
\newblock New model for electron-impact ionization cross sections of molecules.
\newblock {\em The Journal of Chemical Physics}, 104(8):2956--2966, 1996.

\bibitem{Vacuum}
J.~Wolf.
\newblock {Size Matters: The Vacuum System of the KATRIN Neutrino Experiment}.
\newblock {\em Journal of the Vacuum Society of Japan}, 52:278--284, 2009.

\bibitem{NEG}
X.~Luo et~al.
\newblock {KATRIN NEG pumping concept investigation}.
\newblock {\em Vacuum}, 81(6):777--781, 2007.
\newblock {Proceedings of the European Vacuum Conference (EVC-9)}.

\bibitem{Susanne}
S.~Mertens et~al.
\newblock {Background due to stored electrons following nuclear decays in the
  KATRIN spectrometers and its impact on the neutrino mass sensitivity}.
\newblock 2012.
\newblock (Submitted to Astroparticle Physics).

\bibitem{Belesev}
A.I. Belesev et~al.
\newblock Results of the troitsk experiment on the search for the electron
  antineutrino rest mass in tritium beta-decay.
\newblock {\em Physics Letters B}, 350(2):263 -- 272, 1995.

\bibitem{Fraenkle}
F.M. Fr{\"a}nkle et~al.
\newblock {Radon induced background processes in the KATRIN pre-spectrometer}.
\newblock {\em Astroparticle Physics}, 35(3):128--134, 2011.

\bibitem{Nancy}
N.~Wandkowsky et~al.
\newblock {Simulation of radon decays and trapped electrons in the KATRIN
  pre-spectrometer}.
\newblock 2012.
\newblock {(To be published)}.

\bibitem{Holt}
E.~H. Holt and R.~E. Haskell.
\newblock {\em {Foundations of Plasma Dynamics}}.
\newblock {MacMillan, New York}, 1965.

\bibitem{Geller}
R.~Geller.
\newblock {\em Electron Cyclotron Resonance Ion Sources and ECR Plasmas}.
\newblock Inst of Physics Pub, 1996.

\bibitem{Mainz1}
C.~Kraus et~al.
\newblock {Final results from phase II of the Mainz neutrino mass search in
  tritium}.
\newblock {\em The European Physical Journal C - Particles and Fields},
  40:447--468, 2005.

\bibitem{WeinheimerECR}
Ch. Weinheimer et~al.
\newblock High precision measurement of the tritium beta spectrum near its
  endpoint and upper limit on the neutrino mass.
\newblock {\em Physics Letters B}, 460(1–2):219 -- 226, 1999.

\bibitem{Rn220ChargeDist}
S.~Szucs and J.~M. Delfosse.
\newblock {Charge Spectrum of Recoiling $^{216}\mathrm{Po}$ in the
  $\alpha{}$-Decay of $^{220}\mathrm{Rn}$}.
\newblock {\em Phys. Rev. Lett.}, 15:163--165, Jul 1965.

\bibitem{ConversionDataRn220}
S.-C. Wu.
\newblock {Nuclear Data Sheets for A = 216}.
\newblock {\em Nuclear Data Sheets}, 108(5):1057--1092, 2007.

\bibitem{ConversionDataRn219}
E.~Browne.
\newblock {Nuclear Data Sheets for A = 215, 219, 223, 227, 231}.
\newblock {\em Nuclear Data Sheets}, 93(4):763--1061, 2001.

\bibitem{ShakeOff}
M.~S. Freedman.
\newblock {Ionization by Nuclear Transitions}.
\newblock {\em Conference: Summer course in atomic physics, Carry-le-Rouet,
  France, 31 Aug 1975; Other Information: Orig. Receipt Date: 30-JUN-76},
  page~18, Jan 1975.

\bibitem{KShakeOffRadon}
M.~S. Rapaport et~al.
\newblock {$K$-shell electron shake-off accompanying alpha decay}.
\newblock {\em Phys. Rev. C}, 11:1740--1745, May 1975.

\bibitem{ShellReorganization}
J.~S. Hansen.
\newblock {Internal ionization during alpha decay: A new theoretical approach}.
\newblock {\em Phys. Rev. A}, 9:40--43, Jan 1974.

\bibitem{LMShakeOffRadon}
M.~S. Rapaport et~al.
\newblock {$M$- and $L$-shell electron shake-off accompanying alpha decay}.
\newblock {\em Phys. Rev. C}, 11:1746--1754, May 1975.

\bibitem{Kassiopeia}
{\em The Comprehensive Guide to Kassiopeia}.
\newblock Internal KATRIN document.

\bibitem{Kassiopeia2}
D.~Furse et~al.
\newblock {KASSIOPEIA - the simulation package for the KATRIN experiment}.
\newblock {(To be published)}.

\bibitem{FlorianDoktor}
{F. Fr{\"a}nkle}.
\newblock {\em {Background Investigations of the KATRIN Pre-Spectrometer}}.
\newblock PhD thesis, Karlsruhe Institute of Technology (KIT), 2010.

\bibitem{SusanneDoktor}
{S. Mertens}.
\newblock {\em {Study of Background Processes in the Electrostatic
  Spectrometers of the KATRIN Experiment}}.
\newblock PhD thesis, Karlsruhe Institute of Technology (KIT), 2012.

\bibitem{Prall}
M.~Prall et~al.
\newblock {The KATRIN Pre-Spectrometer at reduced Filter Energy}.
\newblock (Accepted for publication in New Journal of Physics).

\bibitem{Mainz2}
J.~Bonn et~al.
\newblock The mainz neutrino mass experiment.
\newblock {\em Nuclear Physics B - Proceedings Supplements}, 91(1-3):273--279,
  2001.

\bibitem{Troitsk1}
V.~M. Lobashev et~al.
\newblock Direct search for mass of neutrino and anomaly in the tritium
  beta-spectrum.
\newblock {\em Physics Letters B}, 460(1-2):227--235, 1999.

\bibitem{Troitsk2}
V.~M. Lobashev et~al.
\newblock Direct search for neutrino mass and anomaly in the tritium
  beta-spectrum: Status of troitsk neutrino mass experiment.
\newblock {\em Nuclear Physics B - Proceedings Supplements}, 91(1-3):280--286,
  2001.

\bibitem{RungeKutta}
J.~H. Verner.
\newblock Explicit runge--kutta methods with estimates of the local truncation
  error.
\newblock {\em SIAM Journal on Numerical Analysis}, 15(4):772--790, 1978.

\bibitem{RungeKutta2}
P.J. Prince and J.~R. Dormand.
\newblock {High order embedded Runge-Kutta formulae}.
\newblock {\em Journal of Computational and Applied Mathematics}, 7(1):67--75,
  1981.

\bibitem{RungeKutta3}
Ch. Tsitouras and S.~N. Papakostas.
\newblock Cheap error estimation for runge--kutta methods.
\newblock {\em SIAM Journal on Scientific Computing}, 20(6):2067--2088, 1999.

\bibitem{FerencEl}
Ferenc Gl{\"u}ck.
\newblock {Axisymmetric electric field calculation with zonal harmonic
  expansion}.
\newblock {\em Progress In Electromagnetics Research B}, 32:319--350, 2011.

\bibitem{FerencMag}
Ferenc Gl{\"u}ck.
\newblock {Axisymmetric magnetic field calculation with zonal harmonic
  expansion}.
\newblock {\em Progress In Electromagnetics Research B}, 32:351--388, 2011.

\bibitem{BEM}
P.~W. Hawkes and E.~Kasper.
\newblock {\em {{P}rinciples of {E}lectron {O}ptics}}, volume~1.
\newblock Academic Press, 1989.

\bibitem{scattering6}
A.~Lewis Ford and J.~C. Browne.
\newblock Elastic scattering of electrons by h2 in the born approximation.
\newblock {\em Chemical Physics Letters}, 20(3):284 -- 290, 1973.

\bibitem{scattering7}
J.~W. Liu.
\newblock {Elastic scattering of fast electrons by $\mathrm{H}_{2}$
  (${}^{1}\Sigma_{g}^{+}$) and $\mathrm{N}_{2}$
  ($X^{1}\Sigma_{\mathrm{g}}^{+}$)}.
\newblock {\em Phys. Rev. A}, 32:1384--1394, Sep 1985.

\bibitem{scattering9}
J.~Komasa and A.~J. Thakkar.
\newblock {Small-angle elastic scattering of high-energy electrons by
  $\mathrm{H}_{2}$, HD, and $\mathrm{D}_{2}$}.
\newblock {\em Phys. Rev. A}, 49:965--968, Feb 1994.

\bibitem{scattering4}
S.~Trajmar et~al.
\newblock {Electron scattering by molecules II. Experimental methods and data}.
\newblock {\em Physics Reports}, 97(5):219 -- 356, 1983.

\bibitem{scattering5}
H.~Tawara et~al.
\newblock Cross sections and related data for electron collisions with hydrogen
  molecules and molecular ions.
\newblock {\em Journal of Physical and Chemical Reference Data},
  19(3):617--636, 1990.

\bibitem{scattering8}
M.~E. Rudd et~al.
\newblock {Doubly differential electron-production cross sections for 200 --
  1500 eV $\mathit{e}^{\mathrm{-}}+\mathrm{H}_{2}$ collisions}.
\newblock {\em Phys. Rev. A}, 47:1866--1873, Mar 1993.

\bibitem{CrossSectionHomePage}
Y.~K. Kim et~al.
\newblock Electron-impact cross sections for ionization and excitation.
\newblock {\em http://www.nist.gov/pml/data/ionization/index.cfm}.
\newblock National Institute of Standards and Technology (NIST) Standard
  Reference Database.

\bibitem{Argon}
E.~Gargioni and B.~Grosswendt.
\newblock {Electron scattering from argon: Data evaluation and consistency}.
\newblock {\em Rev. of Mod. Phys.}, 80, 2008.

\bibitem{ConversionDataKrypton}
S.-C. Wu.
\newblock {Nuclear Data Sheets for A = 83}.
\newblock {\em Nuclear Data Sheets}, 92(4):893--1028, 2001.

\bibitem{Penelope}
J.~Bar{\'o}, J.~Sempau, J.M. Fern{\'a}ndez-Varea, and F.~Salvat.
\newblock {PENELOPE: An algorithm for Monte Carlo simulation of the penetration
  and energy loss of electrons and positrons in matter}.
\newblock {\em Nuclear Instruments and Methods in Physics Research Section B:
  Beam Interactions with Materials and Atoms}, 100(1):31--46, 1995.
\newblock Version 2006.

\bibitem{Zboril}
{M. Zbo\v{r}il for the KATRIN collaboration}.
\newblock {Electron ${}^{83}$Rb/${}^{83\text{m}}$Kr Source for the Energy Scale
  Monitoring in the KATRIN Experiment}.
\newblock {\em AIP Conference Proceedings}, 1417(1):154--158, 2011.

\bibitem{ZborilDoktor}
{M. Zbo\v{r}il}.
\newblock {\em Solid electron sources for the energy scale monitoring in the
  KATRIN experiment}.
\newblock PhD thesis, Westf{\"a}lische Wilhelms-Universit{\"a}t M{\"u}nster,
  2011.

\bibitem{Benjamin}
B.~Leiber et~al.
\newblock {Cosmic muon induced background at the KATRIN experiment}.
\newblock {(To be published)}.

\bibitem{Valerius}
K.~Valerius.
\newblock {The wire electrode system for the KATRIN main spectrometer}.
\newblock {\em Progress in Particle and Nuclear Physics}, 64(2):291--293, 2010.

\end{thebibliography}
\bibliographystyle{unsrt}

\end{document}